\documentclass[aps,letterpaper,twocolumn,preprintnumbers,floatfix,superscriptaddress,nofootinbib,byrevtex,longbibliography]{revtex4-1}
\usepackage{CJK}

\usepackage{amsmath}
\usepackage{amsfonts}
\usepackage{amssymb}
\usepackage{graphicx, rotating}
\usepackage{epstopdf}
\usepackage{epsfig}
\usepackage{latexsym}
\usepackage{graphicx}
\usepackage{color}
\usepackage[dvipsnames]{xcolor}
\usepackage{amsmath,amssymb}
\usepackage{multirow}
\usepackage{hhline}
\usepackage{array,makecell}
\usepackage[utf8]{inputenc}
\usepackage{arydshln}

\usepackage{tikz-feynman}
\tikzfeynmanset{warn luatex=false}

\usepackage{slashed}
\usepackage{hyperref}
\hypersetup{colorlinks, citecolor=bluscuro, linkcolor=bluscuro, urlcolor=bluscuro}
\definecolor{rossos}{cmyk}{0,1,1,0.55}
\definecolor{bluscuro}{rgb}{0.15, 0.2, .85}
\definecolor{bluchiaro}{cmyk}{1,.3,0.,0.1}
\definecolor{verdescuro}{rgb}{0.3,0.8,0.3}

\newcommand{\bs}{\hat s}
\newcommand{\bsl}{s}
\newcommand{\os}{{\hat s^\prime}}

\newcommand{\reef}[1]{(\ref{#1})}

\newcommand{\ha}[2]{H^{#1}_{#2}}
\newcommand{\haf}[2]{h^{#1}_{#2}}
\newcommand{\sfrac}[2]{{#1}/{#2}}

\newcommand{\eq}[1]{Eq.~(\ref{#1})}

\newcommand{\arc}{a}

\newcommand{\nn}{\nonumber}

\newcommand{\be}{\begin{equation}}
\newcommand{\ee}{\end{equation}}          
\newcommand{\bea}{\begin{eqnarray}}
\newcommand{\eea}{\end{eqnarray}}
\newcommand{\bc}{\begin{center}}
	\newcommand{\ec}{\end{center}}

\newcommand{\M}{{\cal M}}
\newcommand{\Mh}{{\hat{\cal M}}}

\newcommand{\N}{{\bar P}}

\newcommand{\cl}{
	\begin{minipage}[h]{0.06\linewidth}
		\begin{tikzpicture}
		\begin{feynman}[small]
		\vertex (x1) at (0,.35);
		\vertex (x2) at (.1,.35); 
		\vertex (i1) at (0,0);
		\vertex (i2) at (.25,0); 
		\vertex (i3) at (.25+.15,0.05){$_{\hat s}$};
		\diagram*{
			(i1) --[thick, half left, looseness = 2.7] (i2),
			(x1) --[white] (x2)
		};
		\end{feynman}
		\end{tikzpicture}
	\end{minipage} 
}

\newcommand{\clinf}{
	\begin{minipage}[h]{0.065\linewidth}
		\begin{tikzpicture}
		\begin{feynman}[small]
		\vertex (x1) at (0,.35);
		\vertex (x2) at (.1,.35); 
		\vertex (i1) at (0,0);
		\vertex (i2) at (.25,0); 
		\vertex (i3) at (.25+.19,0.05){$_\infty$};
		\diagram*{
			(i1) --[thick, half left, looseness = 2.7] (i2),
			(x1) --[white] (x2)
		};
		\end{feynman}
		\end{tikzpicture}
	\end{minipage} 
}

\def\th{\theta}

\begin{document}

\begin{CJK*}{}{}

\widetext

\begin{flushright}
{\small 
CERN-TH-2020-161 \\
Saclay-t20/056 }
\end{flushright}

\title{ Positive Moments 
for Scattering Amplitudes}

\author{Brando Bellazzini}
\affiliation{Institut de Physique Th\'eorique, Universit\'e Paris Saclay, CEA, CNRS, F-91191 Gif-sur-Yvette, France}
\affiliation{CERN,  Theoretical  Physics  Department, Rte  de  Meyrin  385,  CH-1211,  Geneva,  Switzerland}
\author{Joan Elias Mir\'o}
\affiliation{International Centre for Theoretical Physics (ICTP),  Strada Costiera 11, 34135, Trieste, Italy
}
\author{Riccardo Rattazzi}
\affiliation{Theoretical Particle Physics Laboratory (LPTP), Institute of Physics,EPFL, Lausanne, Switzerland}
\author{Marc Riembau}
\affiliation{Theoretical Particle Physics Laboratory (LPTP), Institute of Physics,EPFL, Lausanne, Switzerland}
\affiliation{D\'epartment de Physique Th\'eorique, Universit\'e de Gen\`eve,
24 quai Ernest-Ansermet, 1211 Gen\`eve 4, Switzerland}
\author{Francesco Riva}
\affiliation{D\'epartment de Physique Th\'eorique, Universit\'e de Gen\`eve,
24 quai Ernest-Ansermet, 1211 Gen\`eve 4, Switzerland}

\begin{abstract}
\noindent 
{We find the complete set of conditions satisfied by the forward 2 $\to2$ scattering amplitude in unitary and causal theories.
These are based on an infinite set of energy dependent quantities -- the arcs -- which are dispersively expressed as  moments of a positive measure defined at (arbitrarily) higher energies. 
We identify optimal finite subsets of constraints, suitable to bound Effective Field Theories (EFTs), at any finite order in the energy expansion. 
At tree-level arcs are in one-to-one correspondence with Wilson coefficients.  We establish under which conditions this approximation applies, identifying seemingly viable EFTs where it never does. In all cases, we discuss the range of validity  in both energy and couplings, where the latter have to satisfy two-sided bounds.
We also extend our results to the case of small but finite~$t$. 
A consequence of our study is that EFTs in which the scattering amplitude in some regime grows in energy faster than $E^6$  cannot be UV-completed.}

\end{abstract}

\maketitle

\end{CJK*}
\medskip

{Effective Field Theory (EFT) is the universal framework to describe particle physics on the basis  of the principles of quantum mechanics and  relativity. The EFT construction is nicely  independent of the detailed features of the microphysics lying above reachable energies. Yet unitarity, causality and crossing place robust constraints on the structure of its couplings.
In particular, these constraints take the form of sharp positivity
 bounds   from dispersion relations  for the scattering amplitude both forward~\cite{Pham:1985cr,Ananthanarayan:1994hf,Pennington:1994kc,Adams:2006sv,Bellazzini:2016xrt}, 
and at finite angle~\cite{Vecchi:2007na,Nicolis:2009qm,deRham:2017avq,deRham:2017zjm,NimaHuangTalks}, 
 with a multitude of interesting recent applications e.g.~\cite{Komargodski:2011vj,Camanho:2014apa,Bellazzini:2015cra,Bellazzini:2014waa,Bellazzini:2017fep,Cheung:2016yqr,Luty:2012ww,Distler:2006if,Englert:2019zmt,Bellazzini:2019bzh,Bellazzini:2019xts,Chen:2019qvr,Remmen:2019cyz,Zhang:2020jyn}, in addition to the original studies in the context of the chiral Lagrangian \cite{Pham:1985cr,Pennington:1994kc,Ananthanarayan:1994hf}.

In this article we  extend the positivity conditions on the forward amplitude  to what appears to be a complete set.
This is made possible by: \emph{i)} writing the dispersion relations in terms of suitable  energy dependent quantities in the EFT, the \emph{arcs}, which are  directly related to  the Wilson coefficients at tree level  and capture features of the RG evolution at the quantum level; \emph{ii)} noticing that by  unitarity  the arcs correspond to the sequence of \emph{moments} of a positive measure over a compact interval.
The resulting setup precisely fulfils the hypotheses of Hausdorff's moment problem, whose  known solution provides the set of necessary and sufficient positivity  constraints on the forward amplitude.
We also extend some of  our results beyond the forward limit, using the classic result in Refs.~\cite{Lehmann:1958ita,Martin:1965jj} on the positivity of the $t$ derivatives of the imaginary part of the amplitude.

 Our results are related and partly overlap with those obtained by the geometric approach  put forward in Ref.~\cite{NimaHuangTalks}.  An important difference is that our method  enables  to identify \emph{optimal} constraints that involve  a finite number of arcs/Wilson coefficients only. This is the situation closer to questions of phenomenological interest. 

The constraints we present are most effective in derivatively coupled theories, where the forward amplitude is finite in the massless limit. There, the arcs are {\it single scale} quantities that purely depend on the running couplings, and our constraints directly bound  the RG flow (an aspect  briefly touched in Refs.~\cite{Adams:2006sv,Distler:2006if,Luty:2012ww,Bellazzini:2019xts,Bellazzini:2019bzh}).
One  well formulated hypotheses that can be tested in this context  concerns the existence of  symmetries -- exact, accidental or weakly broken -- which may appear  in the low-energy~EFT.
In particular, the non-linear transformations
\be\label{eqsymm}
\phi(x) \rightarrow \phi(x) +b+b_{\mu_1} x^{\mu_1} +\cdots+ b_{\mu_1\cdots\mu_N}  x^{\mu_1} \cdots x^{\mu_N}  \,,
\ee 
with $b's$ traceless, 
are symmetries in EFTs where interactions have many derivatives and, therefore, deliver soft amplitudes, i.e. amplitudes with a fast energy $E$ growth. The  case $N=0$ is familiar: $U(1)$ Goldstone bosons with $2\to2$ amplitudes $\mathcal{M}\sim E^4$. 
{The same behaviour arises for the EFT of massless particles with spin, like for the Euler-Heisenberg Lagrangian in QED, as well as the theories of Refs.~\cite{Liu:2016idz,Bellazzini:2017bkb,Bellazzini:2018paj}.
The $N=1$ case corresponds to Galileons~\cite{Nicolis:2008in}, with $\mathcal{M}\sim E^6$, and so on. We refer to theories with large exponent in $\mathcal{M}\sim E^{2n}$, $n> 2$, as {\it super-soft}~\cite{Bellazzini:2016xrt,Hinterbichler:2014cwa}.} Similarly, longitudinal polarizations in theories with massive spin-$J$ particles have supersoft amplitudes with~$2n\geq 3J$~\cite{Bellazzini:2019bzh}. 
For example, approximate linear diffeomorphisms   suppress the self-interactions from the Einstein-Hilbert term,  relative to the super-soft linearised (Riemann)$^3$ terms. Such a scenario for gravity is incompatible  with tree-level UV completions~\cite{Camanho:2014apa}.
In this work we will show that supersoftness \emph{in general} cannot emerge, neither by structure nor by accident, from any reasonable, weakly or strongly-coupled  UV completion.

This paper is organized as follows. In section \ref{pa}, focussing on the forward limit, we construct the necessary general dispersion relations and derive optimal bounds
on the arc variables.  In sections ~\ref{sec:forwardlimit} and ~\ref{sec:loops} we apply our results, first assuming 
a weakly coupled UV completion and then lifting this assumption while focussing on the more specific case of an abelian Goldstone boson.  
In section~\ref{sec:bf} we consider a first foray of our methodology beyond the forward limit. Finally in section ~\ref{sec:conc} we summarize our results and offer an outlook.


\section{Arcs and Their  Constraints}
\label{pa}

We  study the  $2\to2$ scattering amplitude ${\cal M}$  of a single  particle of mass $m$ (including the limit $m^2\rightarrow 0$ if it exists).  We discuss first the forward limit $t\to 0$  in theories where it is finite, and define $\M(s)\equiv \lim_{t\rightarrow 0} {\cal M} (s, t)$; for concreteness we focus on the spin-0 case, but our results carry over to arbitrary spin and flavour structure~\cite{Bellazzini:2016xrt}.  Extensions to $t\neq 0$ are postponed to section \ref{sec:bf}.
 We will assume   the following analytic properties of ${\cal M}$:
\begin{enumerate}
\item[$\circ$] Crossing symmetry in $s\leftrightarrow u$  with real analyticity, namely $\M(s)=\M^*(4m^2-s^*)$.
\item[$\circ$] The singularities of $\M(s)$ in the complex $s$-plane consist solely of a unitarity cut at physical energies $s>4m^2$, plus the crossing symmetric one at $s<0$, {see e.g. Ref.~\cite{Martin:1969ina}.}\footnote{\label{ftnt1}Resonances with masses $M^2<4m^2$ below threshold, imply additional poles at $s=M^2,3M^2$. We ignore these for simplicity; their inclusion is straightforward.}
\item[$\circ$]  Unitarity, via the optical theorem, implies~$\text{Im}\M(s)\!>0$.
\item[$\circ$] The amplitude $\M(s)$ is polynomially bounded as $|s|\rightarrow \infty$. In particular,  we assume $\M(s)/s^2 \rightarrow 0$ as $s\rightarrow \infty$. 
In the gapped case, validity of this condition is ensured by  the {Froissart bound~\cite{Froissart:1961ux,Martin:1962rt}}.
\end{enumerate}
Exploiting the analyticity of the amplitude  in the upper half plane, we define  
the arc variables (or simply ``arcs''),
\begin{equation}\label{archdef}
\arc_n(\bs)\equiv \int_{\cl}\frac{d\os}{\pi i}\frac{\Mh(\os)}{\hat s^{\prime 2n+3}}\, ,
\end{equation}
where  $\bs\equiv s-2m^2$ is  the crossing-symmetric variable, $\Mh(\bs)\equiv \M(s)$, and  $\cl$ represents  a counterclockwise semicircular path  in the upper half plane of radius $\bs$, as shown in Fig.~\ref{figex}.

\begin{figure}[t]\centering
\includegraphics[width=0.4\textwidth]{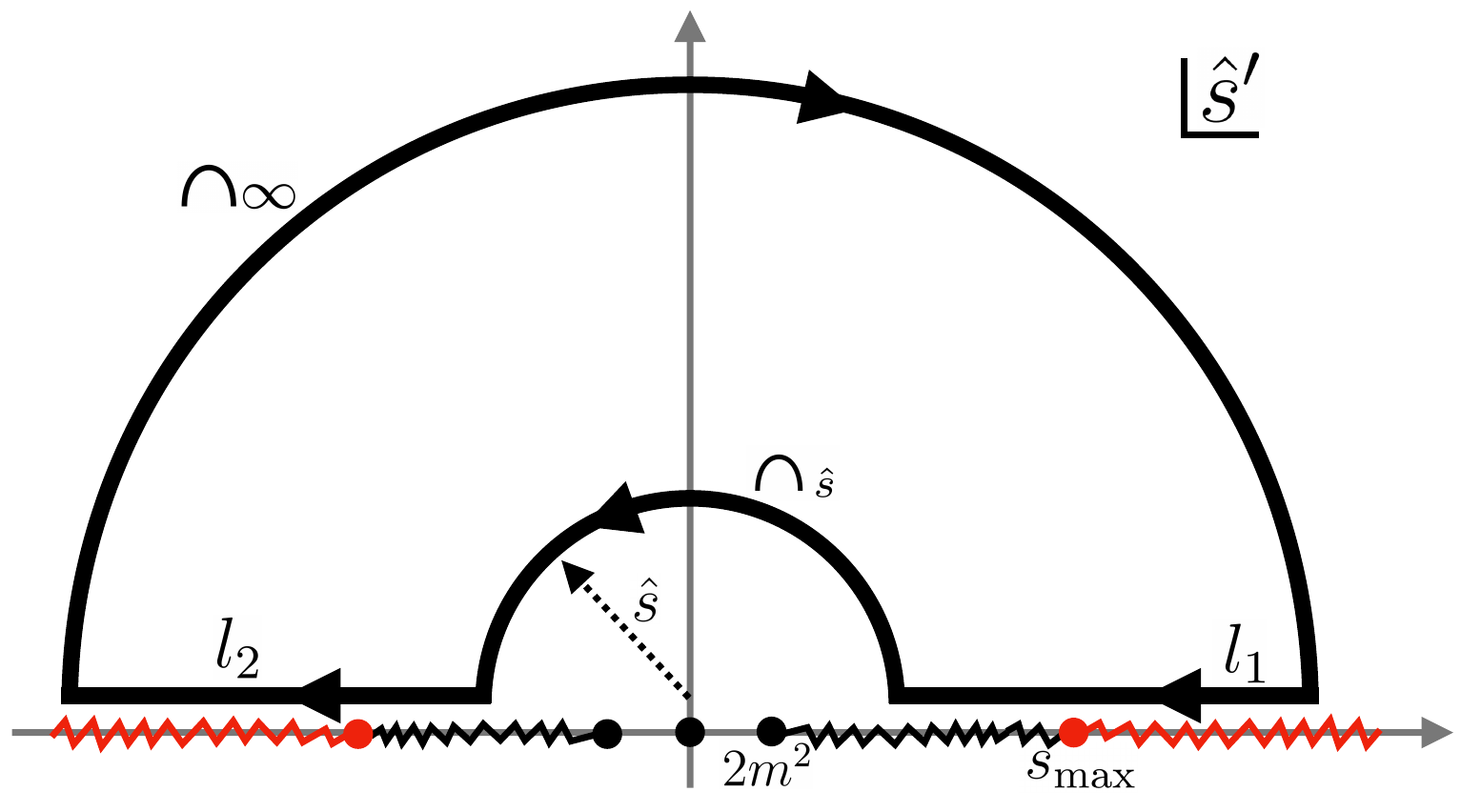}
\caption{{\it The semi-circle contour of \eq{archdef}  in the complex upper $\os$-plane. 
Wiggle lines denote the branch-cuts on the real axis.}  \label{figex}} 
\end{figure}

The Cauchy theorem implies that the integral over the closed contour $C=\cl + \clinf +l_1+l_2$ vanishes. 
Thus, we  deform the integral in \eq{archdef} along   $\cl$  into an integral along $ \clinf  +l_1+l_2 $. 
For integer $n$, we can use crossing symmetry and real analyticity to relate the amplitude  above the lefthand cut (path $l_2$) to the one  above the righthand cut (path $l_1$), 
\be
\Mh(\bs+i\epsilon)=\Mh^*(\bs-i\epsilon)=\Mh^*(-\bs+i\epsilon) \, . \nn
\ee
Due to the  Froissart bound, $\Mh/\bs^2\to0$, the integral along the semicircle of infinite radius  vanishes for ${n\geq0}$. 
Thus, we can write  \eq{archdef}  as
\begin{equation}\label{archint}
\arc_n(\bs)=\frac{2}{\pi}\int_{\bs}^{\infty}d\os\frac{\textrm{Im}\Mh(\os)}{\hat s^{\prime 2n+3}}\,,\quad n\geq0\,.
\end{equation}

{On the one hand, the arcs} as defined via the IR representation \eq{archdef}  
are IR quantities that can be systematically computed as an expansion in powers of $s$ in the domain of validity of the IR EFT $s \ll s_{max}$, with $s_{max}$  the  cutoff.
In the simple case of a tree-level amplitude  in the forward limit,  ${\Mh}(\hat s)= \sum_{n=0} c_{2n} \hat s^{2n}$ and  the arcs match simply to  Wilson coefficients, $\arc_n(\bs) = c_{2n+2}$.
Moreover, the scale dependence of  arcs partly  reflects the EFT RG flow, as we discuss in section \ref{exs}.

{On the other hand}, according to the UV representation in \eq{archint}, arcs receive contributions from all microphysics scales up to the far UV.  
 So,  \eq{archdef} ideally represents something measurable in our low energy experiment, while the representation in \eq{archint} requires knowledge of the theory at all scales. Nevertheless, in this form, \eq{archint} has interesting properties that we now discuss.

\subsection{All  Constraints}\label{sec:constraints}

From \eq{archint}  it follows that the arcs are positive,
\be
\arc_n(\bs ) >0 \, ,  \label{posarch}
\ee
since the imaginary part of the forward amplitude is positive. 
In fact, convoluting in \reef{archint}  a positive   function $F$,  we have 
\begin{equation}
\frac{2}{\pi} \int_{\bs}^{\infty}d\os\frac{\textrm{Im}\Mh(\os)}{\hat s^{\prime 3}} F\left(\frac{\bs}{\os}\right) >0\,.
\end{equation}
For instance, $
F(\bs/\os)= [\bs/\os]^{2n}(1-[\bs/\os]^2) $, with  $\bs < \os$, implies 
\be\label{condii}
\arc_n- \bs^2  \arc_{n+1}  >0 \, . 
\ee
Clearly, to every  function $F$, positive in ${(\bs/\os)^2 \in[0,1]}$, correspond  inequality constraints   relating  arcs of different orders. Characterising the most general such function will allow us to address,

\medskip
\noindent
{\bf Question 1:} What is the complete set of constraints the arcs $a_n$ must satisfy?
\medskip

\noindent 
To answer this question we will relate our problem to the theory of moments. 
The following change of variables 
 \be
x\equiv (\bs/\os)^2\quad d \mu(x)=\frac{ dx }{ \pi }\, \textrm{Im}\, \Mh( \bs / \sqrt{x})    \,   \nonumber
\ee
simplifies the notation and defines a positive measure, so that we can write
\be
\bs^{2n+2} \arc_{n} =\int_0^1     x^n d\mu(x)    \, \, .  \label{an}
\ee
A sequence of dimensionless numbers, defined as in \eq{an} with $d\mu$ positive, is called a \emph{sequence of moments}.
In fact, we have a one-parameter family of sequences because each moment  is a function of  $\bs$. We  comment on the $\bs$ dependence below. 

We introduce   the {discrete derivatives}
\begin{equation}
(\Delta \arc)_n=\bs^2\arc_{n+1}-\arc_{n}\,,
\end{equation}
with higher order differences  defined recursively,  $\Delta^k = \Delta(\Delta^{k-1})$ and $\Delta^0 \arc_n=\arc_n$. For instance, 
$(\Delta^2 \arc)_n = \arc_n - 2\bs^2 \arc_{n+1} + \bs^4\arc_{2 + n}$, etc.
A sequence of moments   necessarily satisfies    
\be
(-1)^k (\Delta^k \arc)_n =\frac{1}{\bs^{2n+2}} \int_0^1 x^n (1-x)^k  d \mu(x) > 0   \nn
\ee
since the functions
\begin{equation}\label{bbernstein}
F(x)=x^n(1-x)^k
\end{equation}
are positive in the whole integration domain. The case $k=1$ corresponds to \eq{condii}.

The converse is   also true, as implied by   the  Hausdorff moment theorem:
 if a sequence satisfies
\be
(-1)^k (\Delta^k \arc)_n> 0  \quad \forall n,k \geq 0, \label{cc}
\ee
then there exists a \emph{unique} measure $d\mu$ such that  \eq{an} is satisfied.\footnote{Arcs probing the theory at finite $s$ are crucial for the mapping to moments on a compact interval (Hausdorff's problem).
Instead, a sequence made of amplitude's residues at $s\to0$ (equivalent to arcs with vanishing radius) see e.g.~\cite{Green:2019tpt}, maps to a non-compact domain (Stieltjes half-moment problem), and the solution is not unique.}
This theorem, following from the fact that the functions in \eq{bbernstein} -- called Bernstein polynomials -- are a basis of all positive functions in $[0,1]$,
provides an \emph{answer to our Question 1}.
 
As mentioned above, both the arcs and their discrete derivatives depend explicitly on  the scale $\bs$, as captured by,
\begin{equation}
\!\!\!\!\!\!\frac{d}{d\bs}[(-1)^k \Delta^k  a_n] \!\!=\!\!\left\{\begin{array}{ll} 
\!\!\!-\frac{2}{\pi}\frac{\mathrm Im \M(\bs +2m^2)}{\bs^{3+2n}}& \!\!(k=0)\\
&\\
\!\!\!2k\bs[(-1)^{k} \Delta^{k-1}  a_{n+1}]& \!\!(k\geq 1)
 \end{array}\right. \label{derivat}
\end{equation}
which is negative for all $k$, because of \eq{cc}.
Therefore, as $\bs$ is increased, the arcs decrease -- proportionally to Im$\M$. This implies that, given an EFT, the constraints \eq{cc} for $k\geq 1$ become more stringent as $s$ increases. Conversely, if the conditions \eq{cc} are satisfied at one scale $\bs$ they are automatically satisfied at smaller scales. This behaviour will play an important role later on, when we discuss constraints on the arcs in specific EFTs.

\subsection{Optimal Bounds for a Finite Set of Arcs}
\label{optimal}

 In practice we often focus on a finite number of arcs. For instance, in a typical EFT only the first few powers of $s$ are phenomenologically interesting -- at tree-level this corresponds  to the first few arcs.
Thus it is natural to ask,
\medskip

\noindent
{\bf Question 2:} 
Considering only a \emph{finite} number $N$ of arcs, what are their optimal constraints?

\medskip
\noindent 
Here \emph{optimal} means the projection of all constraints  on the finite set.
In the language of the previous section, we ask: assuming  the components of $\vec{a}=\{\bs^2 a_0, \dots ,   \bs^{2+2N} a_N\}$ are  moments, what is the subspace  $A(N)\subset \mathbb{R}^N$ on which $\vec a$ takes values?

Of course \eq{cc} still holds and, for $k+n\leq N$, it involves only the first $N$ arcs. 
The constraints with $k+n\geq N$ imply however additional conditions on the subset $n\leq N$. 
 In what follows we will show a simple procedure to extract this information.

Similarly to Question 1 -- that implied finding the most general positive function in $[0,1]$ -- Question~2 requires finding a parametrization for the most general polynomial  $p(x)= \sum_{i=1}^N  \alpha_i x^i $ of finite degree~$\leq N$, positive in $x\in[0,1]$.
Indeed, via \eq{an}, each such $p(x)$ leads to a condition on arcs,
\be
\int_0^1 p(x) d\mu(x) >0 \,\quad \Rightarrow \quad \sum_{i=0}^N \alpha_i a_i >0\,.  \label{constt}
\ee
One can prove that any such polynomial  can be written as   \cite{simon2015real},
\be
p = \sum_J \underbrace{q_{J,1}^2}_\text{type-1}+  \underbrace{xq_{J,2}^2}_\text{type-2} +\underbrace{(1-x) q_{J,3}^2}_\text{type-3}  + \underbrace{x(1-x)q_{J,4}^2 }_\text{type-4}  \label{inst}
\ee
where $q_{J,k}(x)$'s are  non-zero real polynomials -- not necessarily positive --  such that    $p(x)$
is degree $N$, i.e.  $q_{J,k}$ has at most degree $d_k$, with $d_1=\lfloor N/2 \rfloor $, $d_2=d_3=\lfloor (N-1)/2  \rfloor $ and $d_4=\lfloor (N-2)/2  \rfloor $, where~$\lfloor k \rfloor$ is the  integer part of $k\geq0$.
Since \eq{inst} is a sum over positive terms, it is sufficient to discuss them individually. Therefore we drop the index $J$ 
and consider one by one generic polynomials  $q_k$ of each type-$k$ in \eq{inst},  
\be
q_k(x) = \sum_{j=0}^{d_k} \alpha_{kj} x^j  \label{pgen}
\ee
with arbitrary real coefficients $\alpha_{kj}$.

We define the  Hankel matrix $(\ha{\ell}{N})_{ij}= a_{i+j+\ell}$, for $i,j=0,\dots, \lfloor (N-\ell)/2\rfloor$, 
so that $\ha{\ell}{N}$ involves arcs up to $a_{N}$ for $N-\ell$ even and $a_{N-1}$ for $N-\ell$ odd. For instance,
\be
\ha{0}{4}=\ha{0}{5} \equiv\left( 
\begin{array}{ccc} a_0 & a_1 & a_2 \\
a_1  & a_2  & a_3   \\
a_2  &a_3    &a_4 \\
\end{array}
\right)    \, .
\label{poss}
\ee
Now, polynomials of type-1 imply,
\be
\int_0^1 q_1(x)^2  d\mu (x) = \bs^2 \sum_{i,j=0}^{  \lfloor  N/2 \rfloor }  \bs ^{2(i+j)}\alpha_{1i } \, a_{i+j} \,   \alpha_{1j}>0 \, .  \nn
\ee
Since the vector  $\{\alpha_{10}, \alpha_{11}\bs^2, \alpha_{12}\bs^4, \dots \alpha_{1,{N}/{2}}\bs^N\}$ is arbitrary,  the following Hankel matrix  must be positive definite,
\be
\ha{0}{N} \succ 0  \, .  \label{h1c}
\ee
Following similar steps  one finds that the positiveness of  $\int_0^1 x q^2_2(x) d\mu(x)>0$, $\int_0^1 (1-x)q^2_3(x) d\mu(x)>0$ and $\int_0^1 x (1-x)q^2_4(x) d\mu(x)>0$ imply, 
\bea
\ha{1}{N} &\succ   &0 \, ,   \label{h2c} \\
\ha{0}{N-1}- \bs^2\ha{1}{N} & \succ  &0  \label{h3c} \, ,  \\
\ha{1}{N-1} -\bs^2 \ha{2}{N} &\succ & 0 \label{h4c}  \, , 
\eea
respectively. 
We refer to Eqs.~(\ref{h1c}-\ref{h2c}) as \emph{homogeneous} and Eqs.~(\ref{h3c},\ref{h4c}) as \emph{inhomogeneous} constraints.

Since \reef{inst}  is an arbitrary positive polynomial for $x\in[0,1]$,  Eqs.~(\ref{h1c}-\ref{h4c}) represent the optimal constraints,  providing an \emph{answer to Question 2}. 
\begin{figure*}[t]\centering
\includegraphics[height=6cm]{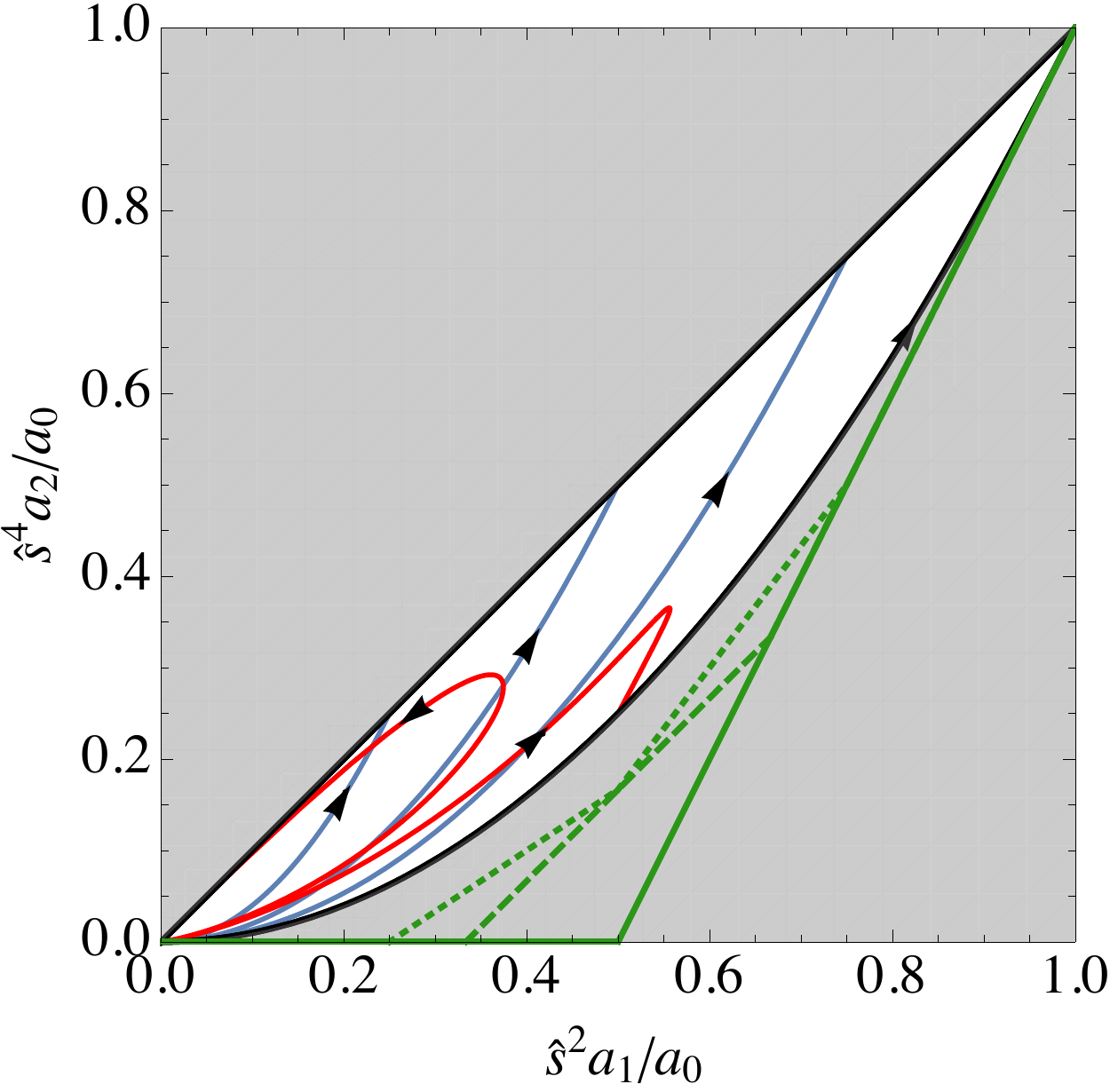}\hspace{1.5cm}
\includegraphics[height=6cm]{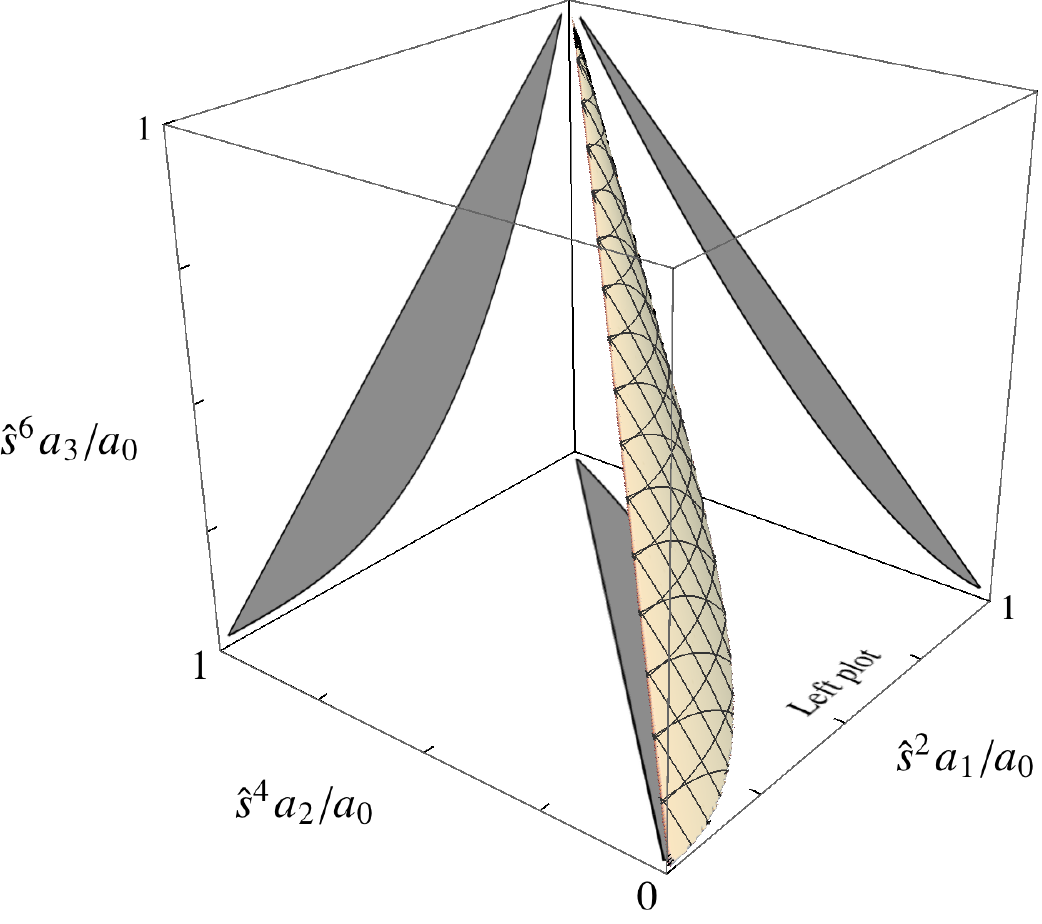}
\caption{\it Allowed regions for the arcs $a_0$, $a_1$, $a_2$ (LEFT) and including $a_3$ (RIGHT), according to Eqs.~(\ref{a2space},\ref{a3space}).
LEFT: For fixed Wilson coefficients, as energy is increased, the theory spans a trajectory in the space of arcs: the blue trajectories (arrows in the direction of increasing $s$) correspond to examples in the weak coupling limit \eq{tree1}, the red trajectories are examples using \eq{arcsgoldstone} at strong coupling   (with the explicit values \eq{explicitgoldstones} and large $g_2$ from \eq{couplings}). {Values of $s^4a_2/a_0$ larger than the green solid/dashed/dotted lines  are excluded by the conditions \eq{cc} from Bernstein polynomials, up to $k+n=2,3,4$ respectively.}
RIGHT: The projections  into  two dimensional planes correspond to optimal bounds when only two coefficients are taken into account (the bottom projection corresponds to the left panel).
The volume of the allowed region is $1/180$ w.r.t. the volume of the unit cube.
 \label{trajs} \label{banana2} \label{banana3}} 
\end{figure*}
For instance, Eqs.~(\ref{h1c}-\ref{h4c})  with $N=2$ define the  $A(2)$ region:
\be
 \left( \begin{array}{cc} a_0 & a_1 \\ a_1 & a_2 \end{array} \right) \succ 0  \, , \ \   a_1 > 0      \, , \ \  a_{0} >\bs ^2a_1   \, , \ \  a_{1} >\bs ^2 a_2 \, .  \label{a2space}
\ee
 This is illustrated in the left panel of Fig.~\ref{banana2}.
The  first constraint in \reef{a2space} implies $a_0a_2>a_1^2$, and  is saturated by the  lowest parabola  in  Fig.~\ref{banana2}; the fourth constraint  in \reef{a2space} is saturated by the upper line, the other constraints imply that the coordinates lie in the interval~$[0,1]$. {For comparison, the green lines show the constraints obtained using \eq{cc} and arcs up to $a_2$ (solid), $a_3$ (dashed), and $a_4$ (dotted), and then projected onto the ($s^2 a_1/a_0$,$s^4a_2/a_0$) plane: clearly these converge to the optimal result A(2).}

Similarly,
evaluating Eqs.~(\ref{h1c}-\ref{h4c})  for $N=3$, we find~$A(3)$: 
\bea
 \left( \begin{array}{cc} a_0 & a_1\\ a_1 & a_2 \end{array} \right)&\succ& 0 \ , \quad \   \left( \begin{array}{cc} a_0-a_1 \bs ^2& a_1-a_2\bs ^2 \\  a_1 -a_2\bs ^2 & a_2-a_3 \bs ^2 \end{array} \right) \succ 0  \, ,  \nonumber \\
  \left( \begin{array}{cc} a_1 & a_2 \\ a_2 & a_3  \end{array} \right) & \succ & 0  \ ,   \quad \  a_{1} >\bs ^2a_2 \, ,\label{a3space}
\eea
which we illustrate in the right panel of Fig.~\ref{banana3}.

Eqs.~(\ref{h1c}-\ref{h4c}) capture how the full constraint on the arc sequence is projected on the first $N$ arcs. Of course we can consider the projection on any subset. In particular for 
 a sequence  $a_k,\cdots, a_N$ that starts at $k\neq 0$, Eqs.~(\ref{h1c}-\ref{h4c}) are generalised by the substitution $\ell\to \ell+k$,  e.g. Eqs.~(\ref{h1c}) becomes $\ha{k}{N} \succ 0$.
\eq{condii} belongs in this class.

{To conclude this section we compare our results to those of Ref.~\cite{NimaHuangTalks}, which -- considering the forward amplitude  -- finds that consistent EFTs must satisfy the set of homogeneous  Hankel matrix positivity constraints, Eqs.~(\ref{h1c}-\ref{h2c}). Indeed, the whole set  of homogeneous constraints (i.e.  for arbitrarily large  $N$) implies the ensemble of constraints in \eq{cc},\footnote{\label{ftntn}{That is because, for $x\in[0,1]$, the Bernstein polynomials in \eq{bbernstein} are arbitrarily well approximated   by a combination of polynomials  of type-1 and type-2 in \eq{inst}, with \emph{arbitrarily large degree}, which precisely correspond to the complete set  of homogeneous constraints  in Eqs.~(\ref{h1c}-\ref{h2c}). For instance $F(x)=1-x$ is reproduced by   $q_1(x)=\sqrt{1-x}=1-x/2-x^2/8-x^2/16-\cdots$ and $q_2=0$, and corresponds to an infinite  Hankel matrix. }} and thus by Hausdorff moment theorem is a necessary and sufficient set.}
However, when only a finite number $N$ of arcs/Wilson coefficients is considered, as it is often the case, our equations Eqs.~(\ref{h1c}-\ref{h4c}) represent the optimal constraints. For instance, if we are interested in the allowed space for three arcs, as in \eq{a2space}, the first two homogenous conditions correspond to simple Hankel determinants and can be easily obtained with the methods of \cite{NimaHuangTalks}. The latter two inhomogeneous conditions, instead, can only be obtained by considering infinite many homogeneous Hankel matrices.

\section{Bounds on Wilson Coefficients}
\label{exs}
{Constraints on  arcs translate, in principle, into constraints on the Lagrangian's Wilson coefficients.
In practice, this translation is complicated by the fact that arcs might receive contributions from (infinitely) many Wilson coefficients. In what follows we discuss under which circumstances this translation is possible:
section \ref {sec:forwardlimit} discusses the tree-level approximation while section~\ref{sec:loops} discusses sizeable quantum effects.}

\subsection{Tree Level}
\label{sec:forwardlimit}
As a first application of our bounds, we focus on the forward limit {in situations where we can consider just the IR tree level amplitude -- we will discuss in the next section under which conditions this approximation holds}. At low energy this takes a polynomial form\footnote{In the single flavour case the amplitude is a function  of $s^2+t^2+u^2$ and $stu$, and  odd powers of $s$ vanish in the forward limit. {See moreover footnote~\ref{ftnt1}.}}
\begin{equation}\label{ampforwardtree}
\Mh(\hat s)=\sum c_d \hat s^d=c_0+c_2 \hat s^2+ c_4\hat s^4+\cdots\,.
\end{equation}
We compute the $\arc_n(\bs)$ using the definition \eq{archdef} and  \eq{ampforwardtree} and find,
 \begin{equation}
\arc_n=\frac{1}{(2n+2)!}\left.\frac{\partial^{2n+2}}{\partial \hat s^{2n+2}}\hat\M(\hat s)\right|_{\hat s=0}=c_{2n+2}\,, \label{tree1}
\end{equation}
independently of {$\bs$}. So the constraints of the previous sections can be read directly  in terms of the coefficients appearing in the amplitude. 
From a practical point of view it is simpler to focus on a limited number of coefficients (rather than the complete series), so that  Eqs.(\ref{h1c}-\ref{h4c})   represent the relevant constraints.
Of these, the homogenous constraints of Eqs.~(\ref{h1c}-\ref{h2c}) { do not depend on $\bs$ and thus represents  properties of  the UV theory that are intrinsic, i.e. independent of the overall scale of the dynamics.} In particular they include \eq{posarch}, which implies that all coefficients $c_n$  be strictly positive~\cite{Adams:2006sv}. 

On the other hand, the energy scale $\bs$ appears explicitly in Eqs.~(\ref{h3c}-\ref{h4c}) (this is translated in the normalisation of Figs.  \ref{banana2}  being $\bs$-dependent). Given an EFT, in the form of a set of $c_n$ satisfying Eqs.~(\ref{h1c}-\ref{h2c}), we can think of Eqs.~(\ref{h3c}-\ref{h4c}) { as defining} the highest possible cutoff $\hat s_{max}$ where new dynamics must modify our EFT amplitude.

For the simple case of the first three coefficients, {in addition to positivity of each of them},   \eq{a2space} and \eq{tree1} imply 
\begin{equation}
 c_2-\bs^2 c_4>0\,,\quad c_4-\bs^2 c_6>0\, , \quad c_2 c_6 > c_4^2\,. \label{3arcstreelevel}
\end{equation}
For fixed values of the Wilson coefficients, as $\bs$  increases, the arcs track
trajectories  in Fig.~\ref{trajs} (blue lines, for different values of $c_n$), that start for $\bs\to 0$ at the origin and evolve along parabolae.
In this case the {cutoff must satisfy } $\bs^2_{max}<\sfrac{c_4}{c_6}$.

The simplest inhomogeneous constraints~${c_2\bs^2-c_n\bs^n>0}$, suffice to rule out any, even approximate, supersoft behaviour, where the tree-level forward amplitude is dominated by the $O(s^{n})$ growth, $n>2$.
First of all, positivity already implies that supersoft symmetries \eq{eqsymm} are never exact: they are always explicitly broken by a  (possibly small) $c_2>0$. 
They could in principle have been appreciable in the regime of energy~$\bs\gtrsim \left({c_{2}}/{c_{n}}\right)^\frac{1}{n-2}$.
{However our bounds forbid that: super-soft theories cannot consistently be UV-completed at weak coupling, as the expansion in $s$ must be strictly decreasing}~\cite{Bellazzini:2019bzh,Englert:2019zmt}. We will see in the next section that the same conclusion holds also beyond the weak coupling approximation.

As a concrete example, consider for instance  interactions of the form $(\partial\partial\phi)^4$, which is   invariant under the Galilean symmetry $\phi\to\phi+b+b_\mu x^\mu$, and giving $\mathcal{M}(s)\sim s^4$. Purely on the basis of symmetries,
these terms could have naturally dominated the more relevant $(\partial\phi)^4$ interactions, which break the Galilean symmetry. {This is however inconsistent with the first inequality in \reef{3arcstreelevel} which forces $(\partial\partial\phi)^4$ to be subdominant.}
\\

\medskip

\noindent
\emph{Faster UV convergence.} 
So far the bounds in this section  do not depend on  $c_0$ in  \reef{ampforwardtree}. 
This changes if we assume that  $\lim_{s\rightarrow \infty}\hat\M(s)\equiv \M_{\infty}$ is finite.
This is the case, for instance, if the theory in the UV is described by a finite number of resonances  in the  tree-level approximation.

For a finite  $\M_{\infty}$  we  can extend the definition  of  arcs in \eq{archdef} to $n\geq -1$.
Then we can repeat an analysis similar to the one that we did in the previous section for the subtracted amplitude $\hat\M-\M_{\infty}$, 
finding that the  arcs define a sequence of positive moments $\{a_{-1},a_0,a_1, \dots\}$.

Note in particular that  $a_{-1}=\hat\M(0)-\M_\infty$ can be regarded as the difference of effective couplings defined by the value of the forward amplitude in the IR and in the UV respectively,  according to the weak coupling intuition. 
Since the first of the  homogeneous conditions on the arcs is now $a_{-1}>0$, the IR forward amplitude is larger than the UV one.  
 Moreover, 
  $a_{-1} a_1-a_0^2>0$ must be satisfied and can be regarded as an upper bound  on $c_2^2/c_4$ if the tree-level approximation of  Eqs.~(\ref{ampforwardtree},\ref{tree1}) holds. 

 An interesting case that we consider in detail in the next section is the theory of a single Goldstone boson for which $c_0=0$. If the UV completion is  perturbative  $-\M_{\infty}\ll16\pi^2$,  
  then we have 
\begin{equation}
0<\frac{c_2^2}{c_4}\leq -\M_\infty \ll 16\pi^2  \, , \label{uppertreebound}
\end{equation}
on the Goldstone theory.~\footnote{\label{footnote6}As example,  consider  a potential $V=\lambda/4 (|\Phi|^2-v^2)^2$ for a canonically normalized complex scalar field $\Phi$. 
The tree-level forward scattering of the Goldstone bosons gives $\M(s)= \left[s^2/(s+m_h^2)-s^2/(s-m_h^2)\right]/(2v^2)$ where $m_h^2=\lambda v^2$. Therefore, $\M_\infty=-\lambda$ and $c_n=\lambda/m_h^{2n}$, so that  $c_2^2/c_4=\lambda$, which saturates the bound~(\ref{uppertreebound}). }
The upper bound on $c_2^2/c_4$ is in agreement with the expectation that  RG running effects on $c_4$ -- that we will see  Eq.~(\ref{explicitgoldstones}) -- are expected to be small in a weakly coupled  theory.

{

\medskip

\noindent
\emph{The boundary.}\label{prefbboundary} {Which} theories saturate Eqs.~(\ref{h1c}-\ref{h4c})?
Hausdorff theorem implies that the integration measure (the imaginary part  of the UV forward amplitude) is \emph{uniquely} determined if \emph{all} arcs are known. If the measure has support on {finitely many points}, then a finite number of arcs suffices to determine the measure uniquely.

Physically, a measure  that consists  of $P$ {\it distinct} delta functions, $\textrm{Im}\Mh(\bs)= \pi/2 \sum_{k=1}^P g_k^2 M_k^2\delta(\bs-M_k^2)$,   is realised in the tree-level approximation when integrating out {heavy particles with  $P$ distinct masses $M_1<M_2< \ldots< M_P$ and   effective squared couplings $g_k^2>0$. This situation  corresponds to~\footnote{For $P\rightarrow \infty$,   finiteness of  $a_0$ requires that $g_k^2/M_k^{4n+4}$  decays sufficiently fast as $k\rightarrow \infty$. Considering a  large but  finite number of particles is therefore  a good approximation.}
\begin{equation}
a_n=\sum_{k=1}^{P} \frac{g_k^2}{M_k^{4n+4}}\,.  \label{treemoment}
\end{equation}
These arcs lie at the boundary of the $A(2P)$ region  (defined by Eqs.~(\ref{h1c}-\ref{h4c}) with $N=2P$), as we now show. 

{
 In the variable $x\in[0,1]$ of section \ref{optimal}, the measure $d\mu(x)$ consists of strictly positive delta functions located at $x_k= \hat s^2/M_k^4$ for $k=1,\dots,P$. In the EFT, i.e. $\hat s<M_1^2$, there are thus $P$ such delta's within~$(0,1)$. 
 Recalling \eq{inst}, Hankel matrices of order $\N$ correspond to the quadratic
forms generated by integrating $x^a(1-x)^b q_{\N-1}(x)^2$ in $d\mu(x)$,
with $q_{\N-1}(x)$ a generic polynomial of order $\N-1$.
This integral gives a vanishing result
only if  all the $\N-1$ zeroes of $q_{\N-1}(x)$  coincide with  the  $P$ delta's in the measure. This can only happen if $\N-1\geq P$, in which
case the corresponding Hankel matrix has order
$> P$:
 Hankel matrices of order $\leq P$ are strictly positive definite, while they are only positive semi-definite if their order is $>P$. 
 Considering then Eqs.~(\ref{h1c}-\ref{h4c}), we  conclude that, for $\bs$ within the EFT, the arcs are in the interior of $A(N)$ for $N<2P$ and at the boundary
of~$A(2P)$.

This implies that -- in weakly coupled theories --  the measurements of the arcs in the IR allows to indirectly count the number $P$ of  resonances in the UV: $P$ is just the smallest $\N$ for which $\det \ha{0}{2\N}=0$.

The same reasoning as above also implies that for $\bs=M_1^2$, just at the edge of the EFT, the arcs are at the boundary of the $A(2P-1)$ region. Indeed, when $\bs$ approaches $M_1^2$ from below, $x_1=\bs^2/M_1^4$
 approaches the edge $x=1$ of the integration region, and the contribution  of the lightest resonance drops out when integrated against the type-3 polynomial $(1-x)q_{\N-1}(x)^2$ of \eq{inst}. Then, for $\N=P$ the zeroes of $q_{\N-1}(x)$ can be chosen to coincide with the location of the remaining $P-1$ heavier resonances, hence $\det \left(\ha{0}{2P-2}-M_1^4 \ha{1}{2P-1}\right)=0$. On the other hand, for $\N<P$,
$q_{\N-1}(x)$ has fewer zeroes than there are resonances. 
This implies that for $N<2P-1$ \eq{h3c} is still strictly satisfied for  $\bs=M_1^2$ and that $\bs>M_1^2$ is needed in order to violate it: $N=2P-1$ arcs allows an optimal estimate of the EFT cut-off $M_1$, while for $N<2P-1$ the estimate is always suboptimal.}

 {
 In fact, the whole  UV spectrum and couplings can be extracted in the IR by determining the roots of the $P$-th order polynomial in $\bs^2$  saturating \eq{h3c}, $\det \left(\ha{0}{2P-2}-\bs^2 \ha{1}{2P-1}\right)=0$. 
 \footnote{According to eq.~(\ref{treemoment}), the $k$-th resonance contributes to the ${ij}$ entry of any Hankel matrix a term proportional to $M_k^{-4(i+j)}$. 
The determinat associated to \eq{h3c} is thus given by sum of terms $\epsilon_{i_1\ldots i_P} M_{k_1}^{-4i_1}M_{k_{P}}^{-4i_{P}}$ (hence fully antisymmetric in the $k_i$) weighed by  the product of~$g_{k_{i}}^2(1-\bs^2/M_{k_i}^4)$. 
 The sum over  $i_k$ therefore vanishes for $s=M_{k^2}$ since only $P-1$ distinct  $1/M_{k_i}^4$ terms appear in the antisymmetric tensor. 
Then, the couplings $g_k^2$ can be extracted by solving \eq{treemoment} in terms of the arcs.  }}

As an example, consider two heavy particles $\varphi_{i=1,2}$ of masses $M_i$ and trilinear vertex $\frac{g_i}{\sqrt{2}}(\partial\pi)^2 \varphi_i/M_i$, matching Eq.~(\ref{treemoment}) -- here $\pi$ is the massless state associated with the $2\to 2$ amplitude. The first 3 arcs $a_{0,1,2}$  populate the bulk of the allowed region  $A(2)$ of \eq{a2space} and Fig.~\ref{trajs}, even for $\bs =M_1$. An estimate of the cutoff using only these 3 arcs, $\bs^2<\sfrac{a_1}{a_2}$, produces values that are always above the true cutoff~$\bs=M_1^2$.
By instead considering 4 arcs  $a_{0,1,2,3}$, the estimated cutoff becomes exact: \eq{h3c} is satisfied for $\bs=M_1^2$. Including 5 arcs, \eq{h1c} is found to be marginally satisfied, i.e. $\det\ha{0}{4}=0$, thus determining the number of states.

}

}

\subsection{Beyond Tree Level}\label{sec:loops}

{Under which circumstances is the  {tree-level formula, $\arc_n=c_{2n+2}$, a good approximation,} and when do the conclusions of the previous section hold?

To answer this, we study an  EFT including RG effects, restricting  for simplicity to the massless case (where $\bs=s$). {Moreover, in order for the forward limit to be well-defined, 
we focus on the case of a {derivatively coupled scalar, {\it i.e.} a Goldstone boson $\phi$ with symmetry  $\phi\to\phi+b$. 
Indeed, the explicit computation that we present below -- up to two loops and $O(s^6)$ --  does  not exhibit any IR divergence.
Moreover, for the Goldstone theory, the tree-level amplitude is finite at $t=0$, and divergences could originate only from   collinear emissions (soft emission does not affect  the total  cross section
 and hence neither the imaginary part of the amplitude). Using collinear factorization (SCET, see e.g. \cite{Becher:2014oda}), it is easy to see that these are finite, as the positive powers of collinear momenta associated with the Goldstone derivative interactions always compensate putative divergences in collinear propagators.  
 We therefore assume that the forward amplitude is well-defined.
 } }
 
 {In the upper half plane, up to $O(s^6)$, the most general Goldstone boson amplitude starts at order $O(s^2)$ and  reads,}
\begin{gather}\label{genF}
\M(s)\!= \!c_2 s^2  +s^4  \left[c_4\!+\!\beta_4 \log(-is)\right]-i\pi s^5 \beta_5/2   \\
+s^6 \left[ c_6+ \beta_6 \log(-is) +\beta_6^\prime \log^2(-is) \right] +O(s^7)\,, \nn 
\end{gather}
 {where  $\log$ is defined in the standard way, with the cut on the negative real semi-axis. Then we have
 $2\log(-is)=\log(s)+\log(-s)$ for Im$s\geq 0$ and  $\log(s)-\log(-s)=i\pi$.  For ease of notation we set the RG scale $\mu=1$, but the generic choice is reinstated through $\log( -is)\to \log(-is/\mu^2)$, $c_n\to c_n(\mu)$.}
 {An explicit calculation in the  Goldstone case gives}
\footnote{By a slight abuse of notation, $c_4$ in \eq{genF} differs from the tree-level amplitude in \eq{ampforwardtree} by a finite one-loop piece: $c_4^{(\ref{genF})}= c_4^{(\ref{ampforwardtree})} + 449c_2^2/(300 (16\pi^2)) $.} 
\bea
\beta_4  &=&  -\frac{7}{10}\frac{c_2^2}{16\pi^2}\ , \quad \beta_5=+\frac{4}{15} \frac{c_2 c_{2,1}}{16\pi^2}\ ,\nn \\[.2cm]
 \beta_6  &=&   -\frac{  83}{70 } \frac{c_4 c_2}{16\pi^2}   - \frac{1}{30}\frac{ c_{2,1}^2}{16\pi^2} - \frac{319}{175}\frac{c_2^3}{(16\pi^2)^2} \ , \ \quad \label{explicitgoldstones} \\[.2cm]
  \beta_6^\prime &=&   \frac{83}{200} \frac{c_2^3}{(16\pi^2)^2} \ ,\nn
\eea
where $c_{2,1}$ is the coefficient of $s^2t$ in the non-forward amplitude.
{At $O(s^6)$ in the energy-expansion}, \eq{genF} does not receive any further correction at any number of loops.
From  \reef{genF}, the first three arcs read,
\begin{eqnarray}
\arc_0\!&=\!&c_2+\!\frac{\bsl^{2}}{2}\beta_4\!+\!\frac{\bsl^3}{3}\beta_{5}\!+\!\frac{\bsl^{4}}{4} \Big(\beta_{6}\! + \!\frac{\beta_6^\prime}{2}(4 \log \bsl\!-\!1)\Big)\!+\!\cdots\,,\nn\\
\arc_1\!&=\!&c_4(\bsl)+\bsl\beta_{5}+\frac{\bsl^{2}}{2}\Big(\beta_6+\beta_6^\prime(2 \log \bsl-1)\Big)+\cdots\,,\nn\\
\arc_2\!&=\!&-\frac{\beta_{4}}{2\bsl^2}-\frac{\beta_{5}}{\bsl}+c_6(\bsl)+\cdots\,, \label{arcsgoldstone}
\end{eqnarray}
where 
\begin{eqnarray}
c_4(\bsl)&\equiv& c_4+ \beta_4 \log \bsl\,\label{rg46}\\
c_6(\bsl)&\equiv& c_6+ \beta_{6}\log \bsl+\beta_6^{\prime}(\log ^2\bsl-\frac{\pi ^2}{12})\,.\nn
\end{eqnarray}
These functions are  RG invariant by construction, and are the natural extensions of $c_{4,6}$ in the interacting theory.

The dots in \eq{arcsgoldstone}
 denote higher powers} of $s$, of which there are a priori infinitely many. If all such contributions were {important}, the EFT would have no predictive power. 
A meaningful EFT exists only under certain assumptions on the convergence of the series. {To this aim, a first unavoidable assumption, is the perturbativity of the dimensionless couplings $g_n^2\equiv c_n (s) s^n$ that control the IR  loop expansion,
\begin{equation}\label{couplings}
g_n^2\equiv c_n (s) s^n\ll (4\pi)^2\,.
\end{equation} 
This condition, {which  we assume throughout this work}, implies that many of the higher order terms that (could) enter \eq{arcsgoldstone} must be small. For instance it implies that $\beta_4 s^2$ and the contribution $\propto s^4 c_4c_2/16\pi^2 $ in $\beta_6s^4$ -- see \reef{explicitgoldstones} -- are  subleading in~$\arc_0$.

 {Nevertheless,  \eq{couplings} does not yet imply the validity of the tree-level approximation  for the arcs.  
Indeed, the corrections to the tree level result $\arc_n=c_{2n+2}$, are controlled by a broader set of 
parameters given by  the ratios}
\begin{equation}\label{relpert}
\frac{\beta_n s^n}{c_m (s)  s^m}\,.
\end{equation}
We will refer to the situations where these parameters are small as {\emph{strong perturbativity}}.
For $n< m$ these parameters grow in the IR and {become  smaller} in the UV, and viceversa for $n>m$: {in the standard RG parlance they  respectively correspond to {\it relevant and irrelevant} deformations away from tree-level.}  
{For $n=m$,  they capture instead the logarithmic} RG running of Wilson coefficients  (they measure the interaction strengths at the cutoff --  see \reef{uppertreebound} for the weakly coupled case). 
For instance, in the sigma-model example of footnote~\ref{footnote6}, $c_n=\lambda/m_h^{2n}$ and $\beta_n\sim (\lambda^2/16\pi^2)/m_h^{2n}$ are characterised by one scale and one coupling so that
the ratios in \eq{relpert} are  of order $(\lambda/16\pi^2)(s/m_h^2)^{n-m}$.
 Then, given $\lambda/16\pi^2\ll 1$ and $s/m_h^2<1$,
 for $n\geq m$ the parameters of \eq{relpert} are always small (strongly perturbative). However, for  $n<m$, they can be larger than unity at small enough energies,
\begin{equation}
s\lesssim m_h^2 \left(\frac{\lambda}{16\pi^2}\right)^{\frac{1}{m-n}}\,.
\end{equation}

In fact,  these quantum effects always dominate the  far  IR  $s\rightarrow 0$, where  the arcs  asymptote to
\begin{equation}\label{arcsIR}
a_0\to c_2\,,\quad a_1\to \beta_4 \log s \,,\quad a_{n\geq2}\to -\frac{\beta_4}{(2n-2)\bsl^{2n-2}}\,.\nn
\end{equation}
 {For $c_2>0$, and  given $\beta_4<0$, as  implied by unitarity within the EFT
\begin{equation}\label{arcsder}
0> -\bsl^{-4}\frac{2}{\pi}\textrm{Im}\M(s)= \beta _4+O(s)\,,
\end{equation}
 these arcs  fulfil all the constraints.}
What happens is that for $\bsl\to0$, the arcs are fully dominated by the IR tail of the spectral density, which is positive and fully determined by the leading term $\propto c_2^2 $ in the $2\to 2$ cross-section. The Hausdorff condition is then trivially satisfied
and, as graphically represented  in Fig. \ref{trajs},  all red trajectories flow to a common attractor as $\bsl\to 0$.

{When higher energies are considered, predictivity is only retained if contributions above a certain finite positive power of $s$ remain negligible, in particular when considering the arcs  \reef{arcsgoldstone}. In view of that,
 we will now discuss two scenarios for omitting higher order terms. We dub them \emph{Simplest EFT}
and \emph{Next-to-Simplest EFT}}. 
 In the first case we assume  that  all irrelevant contributions to the arcs are negligible, namely
 \begin{equation}\label{dica33}
\beta_n s^n\ll c_m(s) s^m  \quad \textrm{for}\,\,n> m \,.
\end{equation}
 This is the standard {situation} in the context of EFTs.  
{Instead, more complex scenarios arise by allowing    irrelevant parameters to sizeably contribute to the arcs}. In the    \emph{Next-to-Simplest EFTs},
we will allow
   \be
 \beta_{n}s^n     \sim  c_m(s) s^m \quad   \textrm{for some}\,\,n> m  \, , \label{nseft}
 \ee
 while all the other $\beta$ coefficients still fulfil \eq{dica33}.

\vspace{5mm}

\noindent
\emph{Simplest EFT.}  
{The IR relevant contribution to the arcs drastically modify the allowed region for the running Wilson coefficients $c_n(s)$}. This is however not apparent in the first two arcs $a_0$ and $a_1$. 
Indeed, for the first arc alone we have ${a_0\simeq c_2}$,  since there are no possible relevant nor marginal perturbations that enter. Therefore the positivity constraint on $a_0$ gives $c_2>0$,
equivalent to the tree-level case. 

Consider now {the first  two arcs. The size of $\beta_4/c_4$ controls the logarithmic running in  $a_1\approx c_4(\bsl)$. The bounds of section \ref{sec:constraints}, in particular the optimal bounds   in Eqs.~(\ref{h1c}-\ref{h4c}), read}
\begin{eqnarray}\label{twoarcs}
&&c_4(\bsl)>0\\
&&{c_2}-c_4(\bsl){\bsl^2}>0\,,\label{twoarcs2}
\end{eqnarray}
in addition to $c_2>0$.
{Written in term of the running coefficient $c_4(s)$, Eqs.~(\ref{twoarcs},\ref{twoarcs2}) have the same form as at tree-level constraints of \eq{3arcstreelevel}}. Notice in particular that  $c_4(s)$ ({the lowest running Wilson coefficient}) cannot be negative. Moreover, the second inequality, together with perturbativity defined in \eq{couplings}, justifies our {neglect} of the term $\propto s^6 c_4^2/16\pi^2c_2 $ in~$\beta_8 s^8/c_2 s^2 $. 
In other words, it  implies that these terms automatically respect  strong perturbativity \reef{dica33}.

{In a causal and unitary theory, violation of any of eqs.~(\ref{twoarcs},\ref{twoarcs2}) must correspond to the  failure of the hypothesis,  \eq{dica33}.  This implies that either the EFT is at its cutoff, or that it is non-standard, in the sense that irrelevant higher derivative terms must be included, as we will discuss below  in an example}. \begin{figure*}[ht]
\begin{center}
\includegraphics[height=5.3cm]{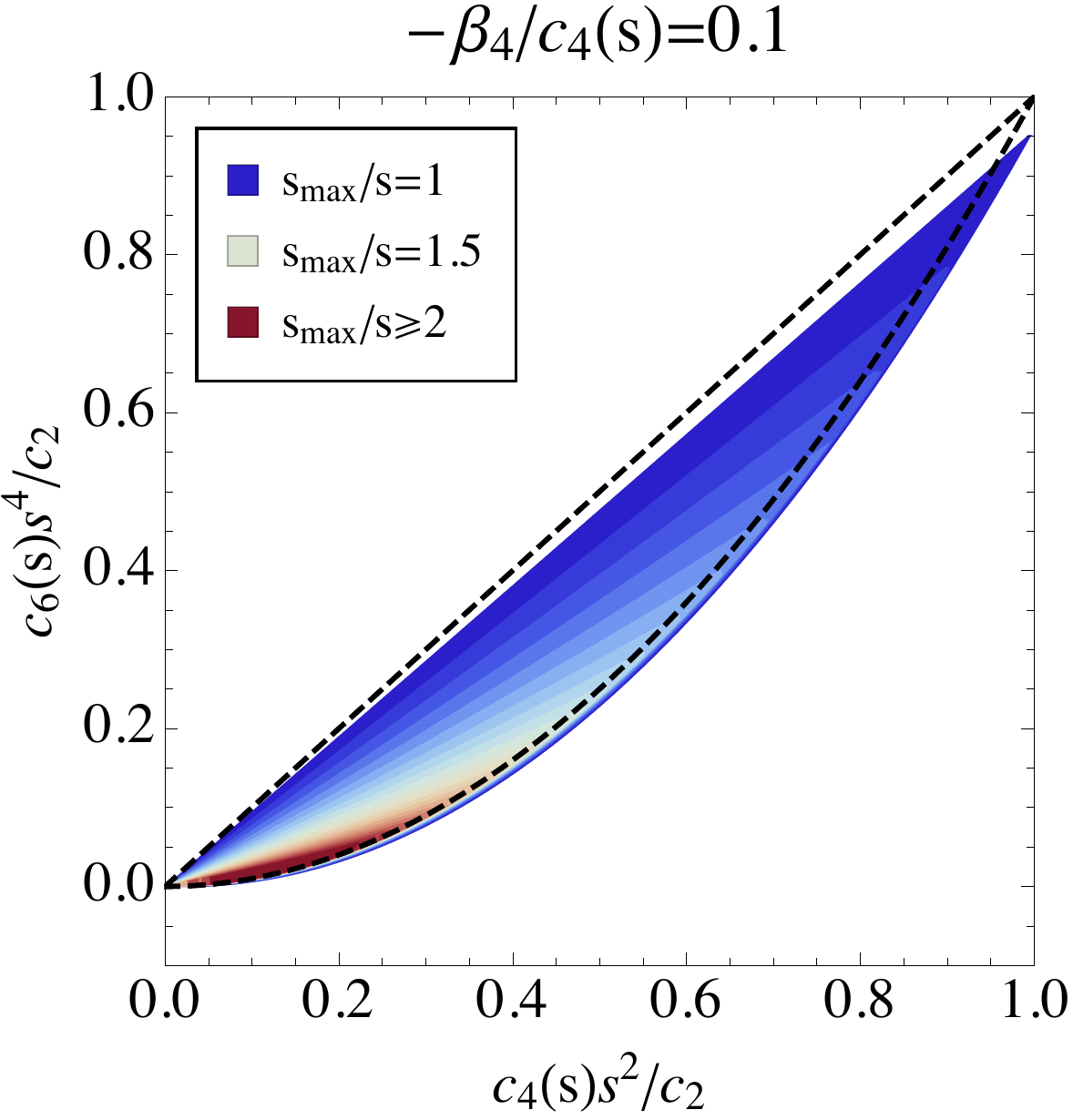}\hspace{2mm}
\includegraphics[height=5.3cm]{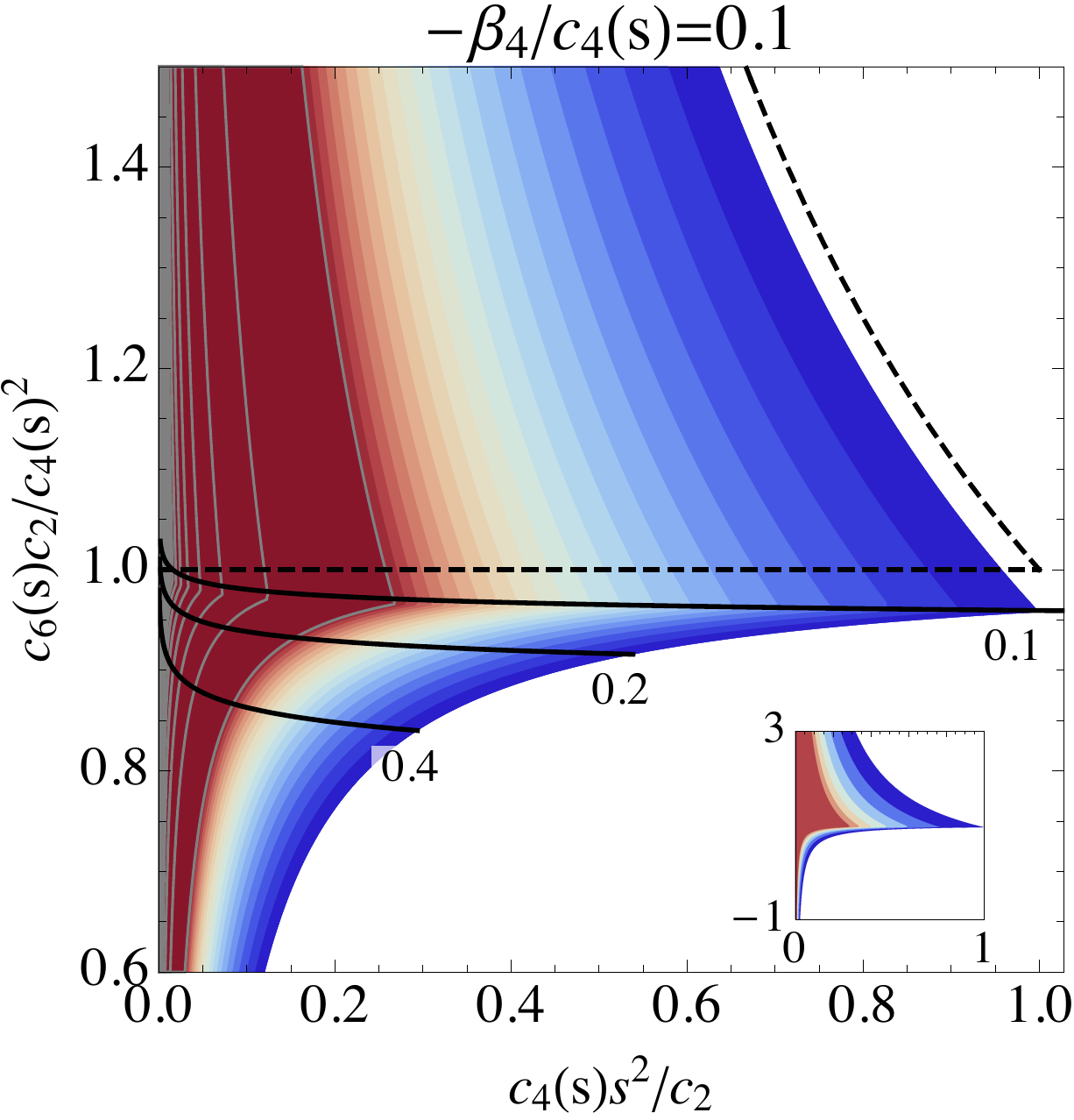}\hspace{2mm}
\includegraphics[height=5.3cm]{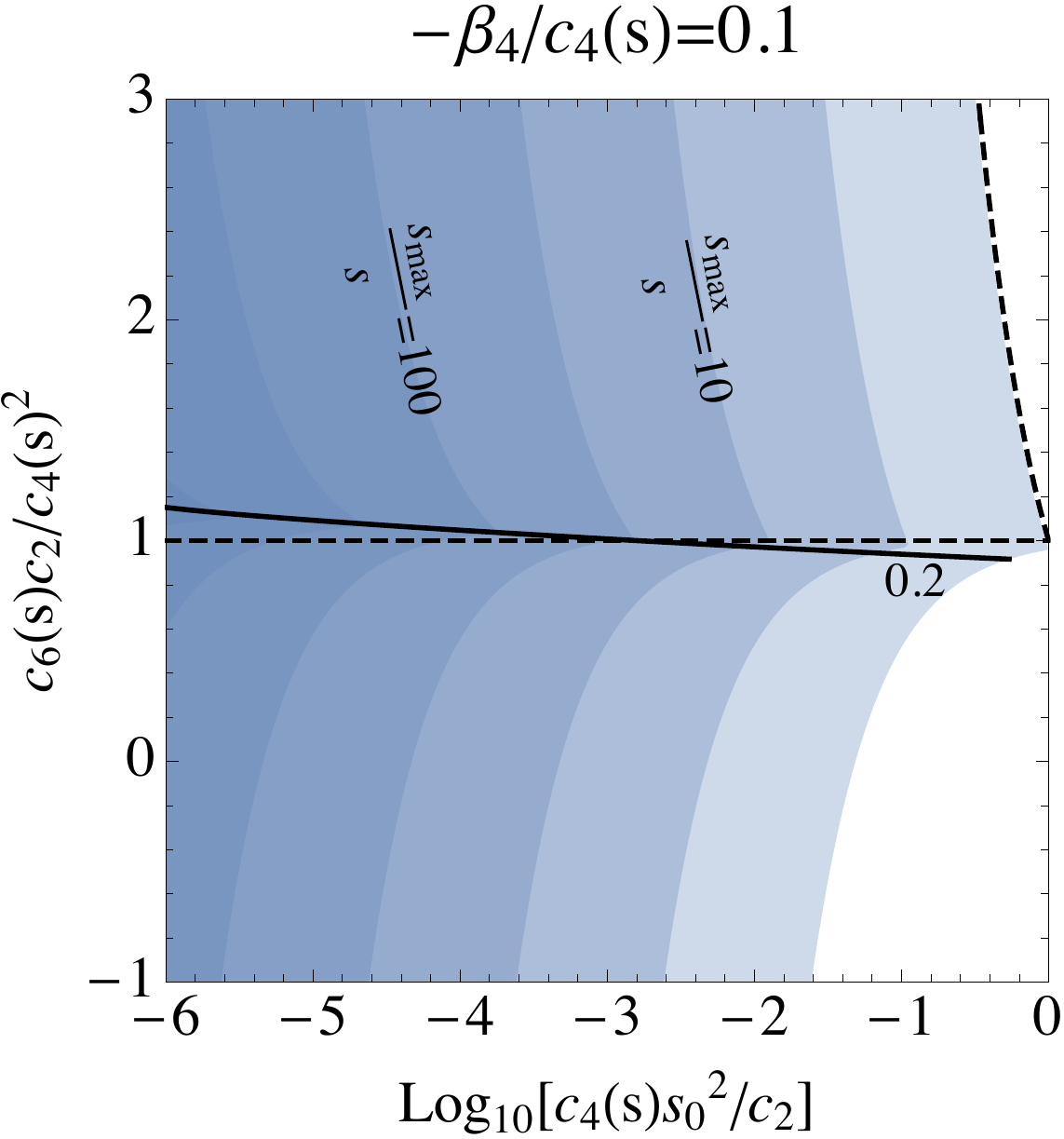}
     \caption{\emph{ In colour the allowed area for (combinations of) Wilson coefficients $c_2$, $c_4(s)$ and $c_6(s)$, evaluated at a scale~$s$. In all panels $\beta_4/c_4(s)=0.1$. 
     Left and center panels: warmer colours denote points where the distance to the cutoff (i.e. the energy $s_{max}$ where  bounds are saturated) is larger; grey contour lines in the central plot  have $s_{max}/s=2,3,4,\ldots$ 
    The black dashed curve denotes the tree-level expectation \eq{3arcstreelevel}. The region above the black lines have  $\beta_4/c_6(s_{max})s_{max}^2<0.1,0.2,0.4$ respectively.  The inset in the center panel shows a wider region of parameter space. 
    Right panel: the same as in center panel but with logarithmic scale. 
 }}\label{f:plot}
     \end{center}
\end{figure*}

\medskip

Starting with the  third arc, the distinction between arcs and running Wilson coefficients becomes apparent. Indeed, the third arc  $a_2\approx -\sfrac{\beta_{4}}{2\bsl^2}+c_6(\bsl)$,\footnote{For simplicity we neglect $c_{2,1}$ -- this is  consistent  because $c_{2,1}$ is not renormalised by the other parameters.  We include sizeable $c_{2,1}$ below, as a case study for the Next to Simplest~EFT.} includes a relevant deformation from its tree-level expectation $\arc_2=c_6$. The bounds read,
\begin{eqnarray}
{c_4(s)}- c_6(\bsl)\bsl^2 > -\frac{\beta_{4}}{2}\,,\label{resultsappr1}\\
c_6(\bsl)-\frac{c_4(s)^2}{c_2}>\frac{\beta_{4}}{2\bsl^2}\,.\label{resultsappr2}
\end{eqnarray}
Given $c_4(\bsl)$ and $c_6(\bsl)$ at an energy $\bsl$, the first expression is stronger and the second is weaker than the tree-level  conditions in \reef{3arcstreelevel}.
The weaker condition implies {that the determinant $c_2c_6(s)-c_4(s)^2$, and indeed even $c_6(s)$,} can be negative. In fact the RG effects in \eq{rg46} alone, already violate the naive tree-level bounds: in the far IR $s\to 0$, RG {evolution leads to  $c_6(s)\to\beta_6^\prime \log^2 s>0$, but  makes the determinant negative,}
\begin{equation}
\det\left(\begin{array}{cc}c_2&c_4(s)\\c_4(s)&c_6(s)\end{array}\right)\to (\beta_4^2-c_2\beta_6^\prime)\log^2 s<0\,,
\end{equation}
where we have used the explicit values for $\beta_4$ and $\beta_6^\prime$ from \eq{explicitgoldstones}. 
{Indeed it is now possible to have consistent EFTs where}  the determinant is so negative that, as energy increases, \eq{resultsappr2} is violated before \eq{resultsappr1}: a radical difference w.r.t. the tree-level approximation, where only the inhomogeneous conditions depend on $s$. This is possible because quantum effects imply that arcs manifestly depend on the energy scale: \eq{derivat} implies that all constraints become more stringent as $\bsl$ increases, so that both homogeneous and inhomogeneous conditions play now a role in defining the theory's regime of validity.

This is illustrated in Fig.~\ref{f:plot}, where coloured regions correspond to Eqs.~(\ref{resultsappr1},\ref{resultsappr2}) and black dashed curves report the naive tree-level expectation.

\medskip

 We can therefore identify two interesting classes of theories. 
First,   theories  that possess a regime in which the parameters in \eq{relpert} are small (in particular $\beta_4/c_4\ll 1$ and $\beta_4/c_6 s^2\ll 1$ in the case of three arcs) and  can be approximated by the tree-level expressions. {Weakly coupled theories where the EFT is obtained by integrating  out massive particles at tree-level (like discussed on page \pageref{prefbboundary}), or  at loop level (like for the Euler-Heisenberg Lagrangian), belong in this class.\footnote{{The discussion in section \ref{sec:forwardlimit} corresponds to the idealized limit where the coupling goes to zero and the tree level regime for  the arcs  extends to arbitrarily small energies. }}}
In theories of this class, it is the inhomogeneous bounds (Eqs.~(\ref{twoarcs2},\ref{resultsappr1}) for three arcs) that are violated first, as $s$ increases. {This state of things is illustrated in Fig.~\ref{f:plot}, where we have chosen $\beta_4/c_4(s)=0.1$, and where  the region above the solid black lines correspond to varying sizes of  $\beta_4/c_6(s_{max})s_{max}^2$, as indicated in the figure.}

 On the other hand,  in theories in which the relevant {perturbation  $\beta_4/(c_6(s) s^2)$ never  becomes negligible}, it is possible for
 $c_2$, $c_4(\bsl)$, $c_6(\bsl)$ to   lie at $O(1)$ outside of the tree-level naive boundary \eq{3arcstreelevel}.
 In the central and right plots of Fig.~\ref{f:plot}, these theories {feature at all energy scales a value of  $c_2 c_6(s)/c_4(s)^2$ significantly below its tree-level bound of 1, and include the option of a negative value, implying  $c_6(\bsl)<0$}. 
 All in all,   even in the \emph{Simplest EFT} scenario, quantum effects open a qualitatively new region in parameter space.
In these theories, as energy increases,  it is  the homogeneous conditions  \eq{resultsappr2} that are saturated first. 
This is illustrated by the  red trajectories in Fig.~\ref{trajs} exiting from the lower parabola.

\vspace{5mm}

\noindent
\emph{Next-to-Simplest EFT.} \label{galpar} 
 We now study the simplest case in which there exist a sizeable effect associated with irrelevant parameters, as in  \eq{nseft}(i.e. a sizeable $\beta_n s^n\sim c_m s^m$ for some $n> m$). The Galileon limit, in which $c_{2,1} \bsl \gg c_2$, is an example of this. 
Since $c_{2,1}$ doesn't enter in the forward amplitude at tree-level, it is not directly bounded by our discussion in section~\ref{sec:forwardlimit}  (in section~\ref{sec:bf} we discuss the amplitude away from the forward limit, but we anticipate that a positive $c_{2,1}$ is unbounded by those arguments).
In this limit, the part of $\beta_6$  involving $c_{2,1}$, which we denote $\hat\beta_6=-c_{2,1}^2/(30(16 \pi^2))$, can be potentially large and depart from strong perturbativity {at sufficiently high energy within the EFT} -- though we still assume (strong) perturbativity for all higher coefficients.

 Considering the first arc, keeping only  the most important contributions in the Galileon limit ($\beta_5$ in \eq{sec:loops} is suppressed w.r.t. $\hat\beta_6$ in this limit), positivity of $a_0\approx  c_2+\sfrac{\hat\beta_6\bsl^4}{4}$ implies
\begin{equation}
 c_2\gtrsim \bsl ^4 \frac{ |\hat\beta_6|}{4} \label{eq:BP} \, ,
\end{equation}
 thus $c_{2,1}$ 
can be at most a loop factor larger than $c_2$, i.e.
$c_{2,1} s^2 \lesssim {8\pi \sqrt{30 c_2}}$,
as already discussed in Refs.~\cite{Bellazzini:2016xrt,Bellazzini:2017fep,Nicolis:2009qm}. 
\eq{eq:BP}  dictates that strong perturbativity between $\beta_6$ and~$c_2$ can be violated only marginally. 

Including also $a_1\approx c_4(\bsl)+\sfrac{\hat\beta_6\bsl^2}{2}$, we find the further conditions
\begin{equation}
-\frac{\hat\beta_6}{2}  \bsl^4\lesssim c_4(\bsl)\bsl^2\lesssim c_2-\bsl ^4 \frac{ \hat\beta_6}{4}\,.\label{secline}
\end{equation}
The first inequality implies that $c_4(\bsl)\bsl^2$  must still be positive. It also implies another bound of the form of \eq{eq:BP}, that can be written explicitly as,
\begin{equation}\label{newgal}
c_{2,1} s \lesssim {8\pi \sqrt{15 c_4(s)}}\,,
\end{equation}
and is stronger than the one implied by \eq{eq:BP}, for~$c_4(s)s^2<2 c_2$.

The second inequality in \eq{secline} shows  that $c_4(s)s^2$ can now be larger than $c_2$. However,  compatibly  with \eq{eq:BP}  it cannot exceed $2c_2$. Therefore the violation of the bound $c_2>c_4(s)s^2$ is only marginal, implying that supersoftness $c_4(s)s^2\gg c_2$ remains forbidden.
 These results are summarised in~Fig.~\ref{figgal}.

If we further include information for the third arc,
$a_2\approx c_6(\bsl)$ (where $c_6(s)$ is dominated by~$\hat\beta_6$ and we neglect the term $\propto\sfrac{\beta_{4}}{\bsl^2}\ll\hat\beta_6$), we have the inhomogeneous bound,{
\begin{equation}
\frac{c_6(\bsl)s^4}{c_2}\left(1+\frac{\hat\beta_6 s^4}{2 c_2}\right)\gtrsim \left(\frac{c_4(\bsl) \bsl^2}{c_2}+\frac{\hat\beta_6 s^4}{4 c_2}\right)^2
\end{equation}
So, a sizeable (negative) $\beta_6$ makes the lower bound on  $\sfrac{c_6(\bsl)s^4}{c_2}$ stronger, and shifts it towards larger values of $\sfrac{c_4(\bsl) \bsl^2}{c_2}$.}
 These effects, however, appear in a relatively uninteresting regime of the theory.
{Indeed, the  quantum effects $\propto \beta_4/s^2$ discussed above were sizeable in the IR and allowed for theories that depart from the tree-level approximation, while still being valid over a relatively large energy regime. Instead, the  effects discussed here, are controlled by the  $s^2 \hat \beta_6 $ term, that is sizeable only at high-energy, when the theory is already close to the cutoff.
}

The couplings $c_{2,1}$, $c_4$ and $c_6$ are all compatible with Galilean symmetry; $c_4$ and $c_6$ are exactly invariant while $c_{2,1}$ is a sort of Wess-Zumino-Witten (WZW) term~{\cite{Nicolis:2008in,Goon:2012dy}}. Instead, $c_2$ violates the symmetry. Our analysis shows that $c_{2,1}$ can be {at most a} loop factor larger than  $c_4$, while the exactly invariant couplings $c_4$ and $c_6$  are directly limited by $c_2$. In some sense, our constraints privilege  the WZW couplings over the exactly symmetric ones.

\begin{figure}[t]\centering
\includegraphics[width=0.4\textwidth]{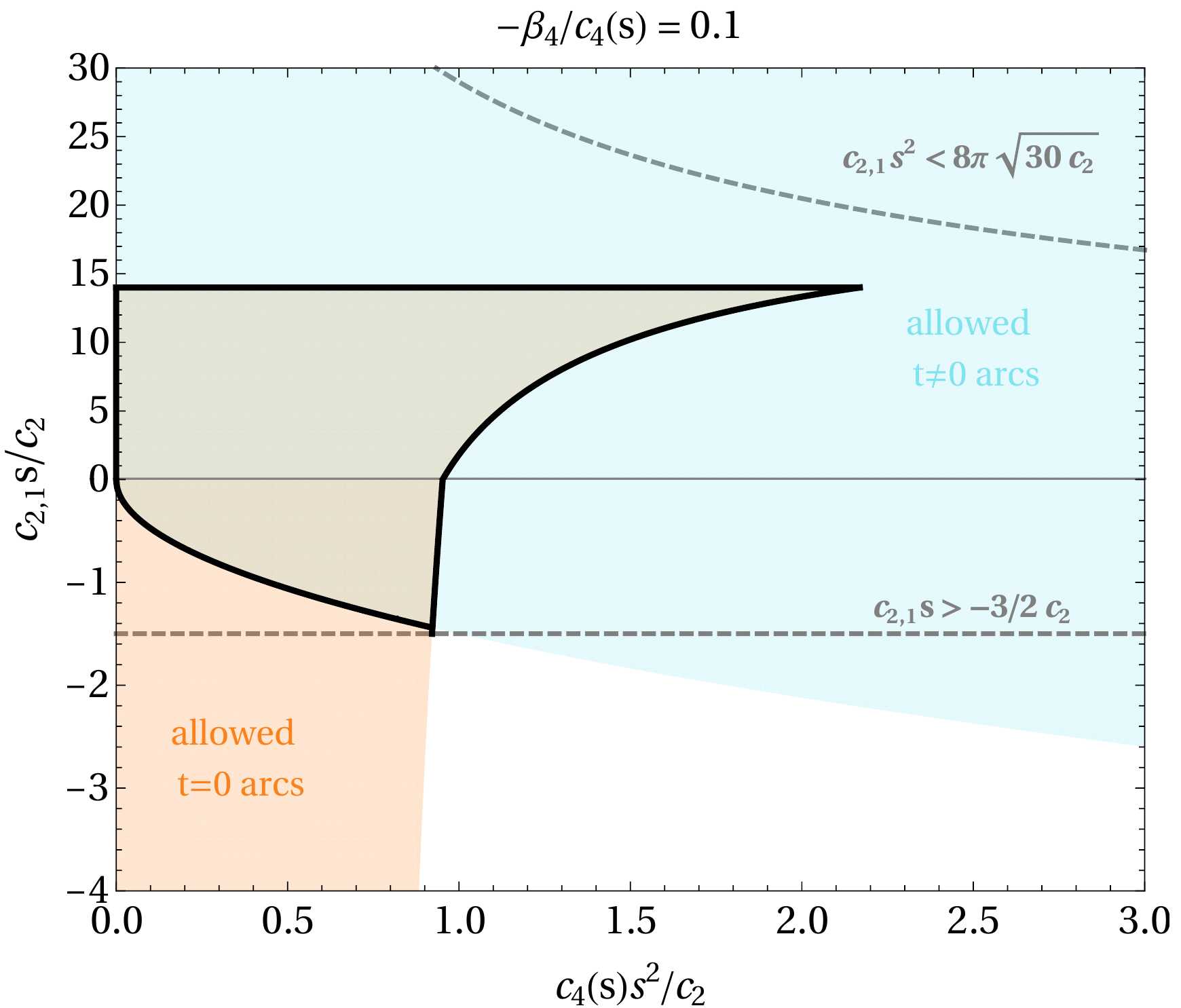}
\caption{\emph{The Galileon case $c_{2,1}\neq 0$, for fixed ${-\beta_4/c_4(s)=0.1}$. Orange Region: allowed from bounds on RGE of the forward $(t=0)$ amplitude (for comparison the upper dashed curve shows the upper bound from Ref.~\cite{Bellazzini:2017fep}). Blue region: allowed by the tree-level $t\neq0$ bounds of section~\ref{sec:bf} (for comparison the lower dashed curve reports the results of Ref.~\cite{deRham:2017avq}). The solid black curve shows the intersection. Notice that upper and lower parts of the plot have different scales.}}
\label{figgal}
\end{figure}


\section{Beyond Forward}\label{sec:bf}

In this section we extend our methodology to the class of amplitudes that are also analytic in $t$ in a finite region around $t=0$. This  trivially
 includes the case  of tree level amplitudes generated by the exchange of massive states, but it also includes the general case of finite mass. Indeed,   for fixed physical $s\geq4m^2$ the amplitude $\M(s,t)$ is analytic for complex $\cos \th \equiv1+\frac{2t}{s-4m^2}$ 
  inside the so called Lehmann ellipse, namely an ellipse with foci at $\cos\th=\pm 1$ \cite{Lehmann:1958ita}, see also e.g.  \cite{Itzykson:1980rh}.

For $t$ finite and real,   $s\leftrightarrow u$ crossing symmetry  and real analyticity   dictate   $\M(s,t)=\M^*(4m^2-s^*-t,t)$.  Further constraints on the $t$ and $s$ dependence are given by the partial  wave  expansion, which diagonalizes the unitarity condition on the S-matrix.  In the physical $s$-channel region the expansion reads
\begin{equation}
\M(s,t)=\sum_{l=0}^\infty P_\ell(\cos\theta)f_\ell(s)  
\end{equation}
while a similar relation also holds in  the $u$-channel. Then, unitarity of the partial waves 
\begin{equation}\label{pospartwav}
\textrm{Im}(f_\ell(s))> 0\quad \forall {\ell}\qquad \mbox{for }s\geq4m^2\, ,
\end{equation}
together with positivity of the Legendre polynomials and their derivatives at $t=0$,
\eq{pospartwav} imply \cite{Martin:1965jj}
\begin{equation}\label{positivedtk}
\left.\partial^k_t  \textrm{Im}\M(s,t)\right|_{t=0}=\sum_{\ell=0}^\infty  \partial^k_t P_\ell(\cos \th)|_{t=0}\textrm{Im}f_\ell(s)>0\,,
\end{equation}
for all $k$ and for $s$  along the $s$-channel cut.  
The above additional positivity conditions can be exploited as we now discuss.

{\medskip

Considering the above properties we   extend the definition of the arcs (\ref{archdef}) to
\begin{equation}\label{archdeft}
\arc_n(\bs,t)\equiv \int_{\cl} \frac{d\os}{\pi i}\frac{\Mh(\os,t)}{ (\hat s^\prime+\frac{t}{2})^{2n+3}}\end{equation}
where we recall $\bs\equiv s-2m^2$, $\Mh(\hat s,t)=\M(s,t)$ and ${\cl}$ is now a contour with radius $\bs +t/2$ centred at~$-t/2$. The condition $\M(s,t)=\M^*(4m^2-s^*-t,t)$ ensures the reality of the arcs. Furthermore 
using that $\Mh(\bs,t)/s^2\to 0$ for  $s\to \infty$ (as dictated by {the analog of the Froissart bound at finite $t$}~\cite{Jin:1964zza}),  the arcs can be expressed by a dispersive integral like in \eq{archint},
\footnote{Since the amplitude is analytic also for $0\leq t \leq 4m^2$ \cite{Martin:1965jj},   $\textrm{Im}\Mh(\bs,t)=\sum_n \partial^n_t\textrm{Im}\M(s,t) |_{t=0}  t^n/n! >0$ is positive there~\cite{Vecchi:2007na,Nicolis:2009qm}, so that  the constraints of section \ref{pa} apply up to replacing $a_n(\bs)\rightarrow a_n(\bs,t)$ for $ 0\leq t\leq 4m^2$. 
In what follows we provide stronger constraints than these, by using that each derivative in Eq.~(\ref{positivedtk}) is separately positive.

{Notice also that it is in principle  possible to build dispersion relations for other combinations that respect real analyticity, such as $\partial_t\Mh-\partial_{\hat s}\Mh/2$. These also lead to \eq{dta}.
}
} 
\begin{equation}\label{archdeftUV}
\arc_n(\bs,t)=\frac{2}{\pi}\int_{\bs}^{\infty} \! d\os\frac{\textrm{Im}\Mh(\os,t)}{ (\hat s^\prime+\frac{t}{2})^{2n+3}} \quad n\geq 0\,.
\end{equation}
At this point we can take $t$-derivatives at $t=0$ and use \eq{positivedtk}  to obtain positivity conditions \begin{eqnarray}\label{dta}
\partial_t \arc_n (\bs,t)|_{t=0}&=&a^{(1)}_n(\bs)- \frac{2n+3}{2}\arc_{n+1/2} (\bs)\,,\\
\partial_t^2 \arc_n (\bs,t)|_{t=0}&=&a^{(2)}_n(\bs)- (2n+3)a^{(1)}_{n+1/2}(\bs)\nn\\&&+ \frac{(2n+3)(2n+4)}{4}\arc_{n+1}(\bs)\,,
\nn
\end{eqnarray}
and so on;  where  we have defined 
\begin{equation}\label{defDa}
a^{(k)}_n(\bs) =\frac{2}{\pi}\int_{\bs}^\infty d\os \frac{\partial^k_t\mathrm{Im}\Mh(\bs,t)|_{t=0}}{\hat s^{\prime 2n+3}}\,,
\end{equation}
and $a^{(0)}_n\equiv a_n$. Notice that while the arcs $a_n(\hat s,t)$ are defined for integer $n$ and can be written purely in terms of IR data according to \eq{archdeft}, the $a^{(k)}_n(\bs)$ of  \reef{dta}, are defined for half-integer $n\geq 0$ and   only through the UV representation in \reef{defDa}.

For every $k$, $\{ \bs^{2n+2} a_n^{(k)}\}$, with half-integer $n\geq 0$, is a series of moments  because the measure in \reef{defDa} is positive according to  \reef{positivedtk}. 
Therefore, they  fullfil versions of  the  positivity constraints, analogous to  Eqs.~(\ref{h1c}-\ref{h4c}), but  including half-integer arcs. 
More precisely, half-integer arcs fulfil  
Eqs.~(\ref{h1c}-\ref{h4c}), recast  for new Hankel matrices $\left(\haf{\ell}{N} \right)_{2i2j}=a_{{i+j}+\ell}$ where $2i,2j=0,1,\ldots, \lfloor N-\ell\rfloor$ (here both $N$ and $\ell$ can be half-integers). The matrices $\haf{\ell}{N} $ contain both integer and half integer arcs,~e.g. 
\be
\haf{0}{1}= \left( 
\begin{array}{cc} a_{0} &a_{1/2}  \\
a_{1/2}  & a_{1} 
\end{array}
\right)  \, .\label{posst}
\ee

Half-integer arcs $\arc_{n+1/2}$ ($n$ integer) and all $\arc^{(k)}_{m}$ ($m$ integer or half-integer for $k>0$)  
are not calculable in the EFT because they are defined  in terms of UV integrals \reef{defDa}, and have no IR counterpart like \reef{archdeft}.
On the contrary, $\arc_n$ and $\partial_t \arc_n$, with integer $n$, can be computed directly within the IR. Our goal is therefore to understand the constraints on  $a_n$ and $\partial_t a_n$ for arbitrary values of $\arc^{(k)}_{m}$ and $\arc_{n+1/2}$, compatible with their being moments. 
In  appendix \ref{sec:appendixt}, we outline an analytic procedure to do so, from which the  bounds below are derived, but that applies in principle to all $N$. For more $t$-derivatives this procedure becomes cumbersome: in Ref.~\cite{toappearLP} we propose a numerical technique, based on semi-definite programming, to extract the bounds efficiently.

}

For instance, for $N=0$  we find the constraint on  arcs $t$-derivatives  to be $\partial_t a_0 >-\frac{3}{2}\frac{a_0}{\bs}$, a condition that appears in different form already in Ref.~\cite{deRham:2017avq}.

For $N=1$ our conditions  constrain the space  $\{\partial_t a_0,\partial_t a_1\}$ as,
\begin{eqnarray}
&&\partial_t a_0> -\frac{3}{2}\sqrt{a_0 a_1}\,,\qquad \partial_t a_1>-\frac{5}{2} \, \frac{a_1}{\bs}\,,\nn\\
&&\partial_t a_0-s^2 \partial_t a_1>\frac{5}{2}\sqrt{\frac{a_1}{a_0}}(\bs^2 a_1 -\frac{3}{5}a_0)\,, \label{finitetsoftkiller}
\end{eqnarray}
which we illustrate in Fig.~\ref{figarcsder}  for negative ${\partial_t a_0}$ (for ${\partial_t a_0>0}$, the allowed region is unbounded: the distance between the upper and lower boundaries of the projection on the {$\{\bs^2a_1/a_0,\bs^3\partial_t a_1/a_0\}$ plane increases with $\bs\partial_t a_0$).
These conditions are more stringent than just $\partial_t a_0 >-\frac{3}{2}\frac{a_0}{\bs}$ (which corresponds to the  left boundary of the box in Fig.~\ref{figarcsder}), as shown also in~Fig.~\ref{figgal}.
\begin{figure}[t]\centering
\includegraphics[width=0.4\textwidth]{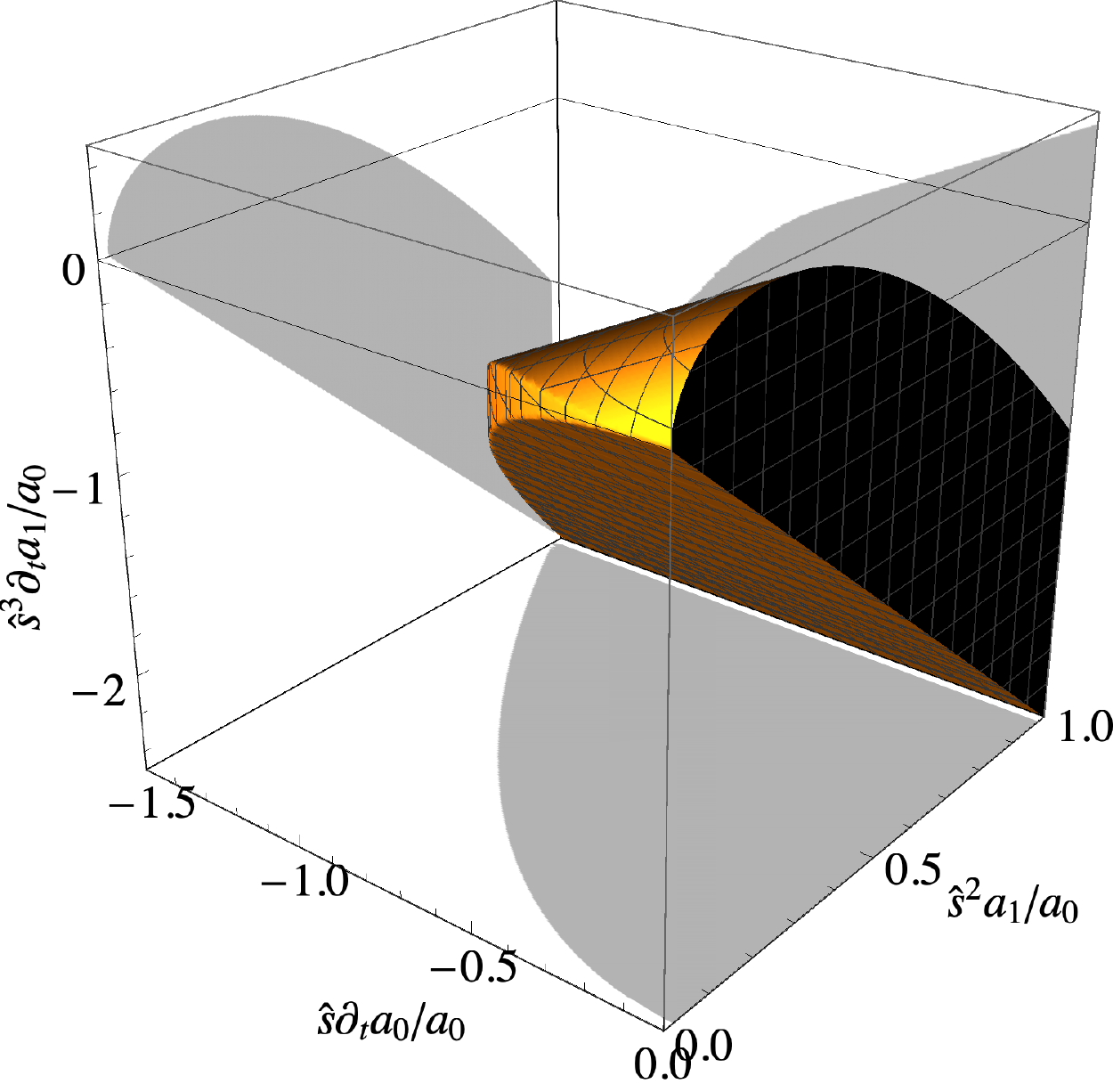}
\caption{\emph{Allowed region in the space of arcs and their first $t$-derivative  for  ${\partial_t a_0<0}$, according to \eq{finitetsoftkiller}; 2D projections in grey. At tree level we have $\arc_0=c_2$, $a_1=c_4$, $\partial_t\arc_0=c_{2,1}$ and $\partial_t\arc_1=c_{4,1}$. }}
\label{figarcsder}
\end{figure}

\vspace{0.5cm}

We illustrate an application of these bounds to the Wilson coefficients at  tree-level}\footnote{At one loop, for a $U(1)$ Goldstone boson (i.e. with $c_0=0$),
\begin{equation*}
\left.\partial_t\M\right|_{\substack{t\to0\\ m\to0}}\!\!= c_{2,1}s^2-\frac{i41 \pi c_2^2}{60(16\pi^2)} s^3 +\left(\!\!c_{4,1}\!-\!\frac{11 c_2 c_{2,1}}{15 (16\pi^2)}\log(-is)\!\right)\!s^4
\end{equation*}
with $t\rightarrow 0$ before taking $m\rightarrow 0$,  is finite. Therefore, for statements up to $O(s^4t)$, the use of the tree-level expressions is justified. We postpone a more refined discussion of these loop effects to Ref.~\cite{toappearLP}.}
\begin{equation}\label{treeansatzfinitet}
\Mh(\hat s, t)=\sum_{n+m>0} c_{n,m} \hat s^n t^m \,.
\end{equation}
\eq{finitetsoftkiller} reads at $t=0$,
\begin{eqnarray}
\label{explicittexample}
c_{2,1}> -\frac{3}{2}\sqrt{c_4 c_2}\,\quad\quad \bs\, c_{4,1}>-\frac{5}{2}c_4 \nn\\
c_{2,1}-\bs^2 c_{4,1}>\frac{5}{2}\sqrt{\frac{c_4}{c_2}}(\bs^2 c_4 -\frac{3}{5}c_2) \label{explicittexample2}
\end{eqnarray}
where we have identified $c_n\equiv c_{n,0}$ to match  the notation of the previous sections.

\eq{explicittexample} implies that $c_{2,1}$  can be negative but not arbitrarily so, as it is limited in magnitude by $\sqrt{c_4 c_2}$. 

To further illustrate the constraining power of \eq{explicittexample2}, we can for instance use it to test the consistency of  a theory  where  the amplitude is  dominated by $\propto E^{10}$ terms for sufficiently large $E$ within the EFT domain of validity. By Lorentz invariance the only option is $\M\simeq stu(s^2+t^2+u^2)$}, corresponding to a $c_{4,1}s^4 t$ term dominating $c_{2}s^2$,  $c_{2,1}s^2t$ and $c_{4}s^4$.  This hierarchy appears natural, since it is protected by one of the approximate symmetries in \eq{eqsymm}, see~\cite{Hinterbichler:2014cwa}.
 Such example is not constrained by our tree-level  forward bounds in section \ref{sec:forwardlimit}, simply because the largest contribution to the amplitude vanishes at~$t=0$.
However,  \eq{explicittexample} provides upper and lower bounds  on $c_{4,1}\bs^5$,  controlled by more relevant Wilson coefficients,  see~Fig.~\ref{figarcsder} and its caption. 
In the physical region~$|t|<s$,  these bounds imply that $c_{4,1}s^4 t$ can never dominate the   terms with lower powers of ~$E$. In other words, supersoft amplitudes that vanish in the forward  limit and have more powers of energy than~$c_{2,1}s^2 t$, are  excluded by our bounds. 

The same arguments lead to constraints on higher $t$-derivatives of arcs. The structure is always the same:  $\partial_t^k a_0$ are bound from below, but can be arbitrarily large when positive. Instead $\partial_t^k a_n$, $n>0$ are bound from below and from above. 

At tree level, this means that $c_{n,k}$, with  $n>2$ for any $k$, fulfil two-sided bounds.
Moreover, in the single flavour case considered here, crossing symmetry  implies that $c_{2,k}$ with $k\geq 2$ are related to $c_{2+k,0}$ (for $k$ even) or $c_{1+k,1}$ (for $k$ odd), which are already bounded from above  from our constraints (grey area in table~\ref{tablecolor}). For instance, the amplitude $\M(s,t)\propto (s^2+t^2+u^2)^2$ implies $c_{2,2}=3c_4$, and $c_4$ is bounded from above.
In other words, only $c_2$ and $c_{2,1}$  are unbounded from above by  tree-level arguments; all other coefficients are instead bounded from below and above.
We illustrate this in table~\ref{tablecolor}.

\begin{table}\centering
\includegraphics[width=0.3\textwidth]{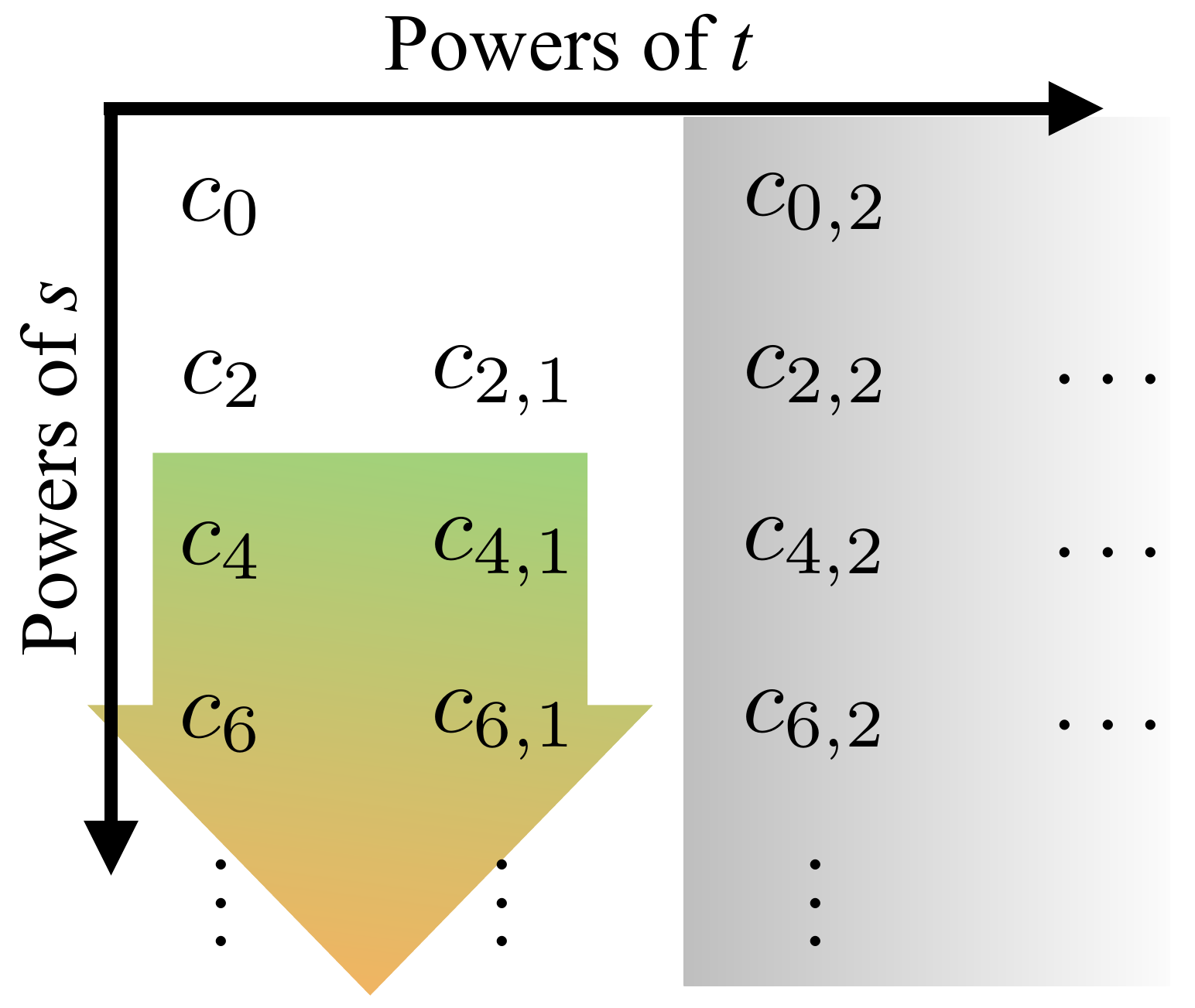}
\caption{\footnotesize\emph{{Schematic summary of  tree-level results. The grey area encompasses coefficients which are not independent because of crossing symmetry in the single-flavour case. 
 $c_0$ is unconstrained, $c_2$ is constrained to be positive, $c_{2,1}$ is constrained to be larger than a combination of the other coefficients. All other coefficients are bounded below and above.}}}\label{tablecolor}
\end{table}%

As pointed out in Ref.~\cite{NimaHuangTalks}, \eq{positivedtk} contains more information than just positivity, which is what we have exploited so far. Indeed, the $\partial^k_t \textrm{Im}\M(s,t) |_{t=0}$ (and their integrals $a^{(k)}_n(\bs)$) are not merely positive, but are the sum of  positive parameters with known coefficients:  $ (s-4m^2)^k\partial^k_t P_\ell(\cos \th)|_{t=0}=\sfrac{(\ell+k)!}{(\ell-k)!k!}$.
This implies more bounds involving at least three $a^{(k)}_n(\bs)$, with $k+n=constant$~\cite{NimaHuangTalks}. Because they involve at least three different $t$-derivatives, and because they are saturated by the crossing symmetry condition in the simplest cases, these bounds didn't play a role in our discussion of supersoftness; for a relevant application see \cite{Huang:2020nqy}. In Ref.~\cite{toappearLP} we will show how these bounds emerge in the language of moments and discuss quantitatively the impact of IR divergences.
}

\section{Summary and Outlook}\label{sec:conc}

In this paper we  introduced a set of energy dependent quantities, the arcs $a_n(s)$ of \eq{archdef}, that
  conveniently  encode the constraints of causality, unitarity and crossing on the forward $2\to 2$ scattering amplitude. A dispersion relation, \eq{archint}, allows to express the arcs as the moments of a positive measure  in the range $[s,\infty)$.
 Hausdorff's moment theorem then establishes the set of necessary and sufficient conditions the arcs must satisfy, given the positivity of the measure. Both the arcs and the constraints are infinite sets. However we derive the projection of the full set of constraints on the subsets of the lowest arcs,  ${a_0,\dots, a_N}$ for any $N$. These are expressed by  Eqs.~(\ref{h1c}-\ref{h4c}) and fall into two classes,  in-homogeneous and homogeneous, 
 according to their explicit, or only implicit, dependence on $s$. These projections are interesting within EFTs, because the lowest $a_n$'s encode the effects of the correspondingly lowest Wilson coefficients.

Our result is particularly relevant to determine the acceptable range of validity  of  EFTs in both couplings and energy. Concerning the latter and as implied by \eq{derivat}, the energy dependence of the arcs leads to stronger
bounds on the EFT parameters as the energy is increased. Satisfaction of the positivity constraints at a certain $s$, automatically implies satisfaction at lower $s$ but not  at higher $s$. When the  parameters of a given EFT violate the constraints above a certain scale, the only option to respect positivity is the breakdown of the  EFT description at or below that scale.

In the idealized limit where the tree level approximation holds exactly, the arc series is in one to one correspondence with the series of Wilson coefficients in the expansion of the forward amplitude. 
The energy dependence of the arcs arises at the quantum level  from  two sources: the  RG evolution  of the Wilson coefficients and collinear radiation from the initial state. The latter  induce effects that typically go like powers of $\ln s/m^2$, and thus diverge in the massless limit,
corresponding to the IR divergence of the total cross section. There are however situations, in particular when the interactions are purely derivative, where these IR divergences are absent and the arcs energy dependence is purely controlled by RG evolution. It is this simpler situation that we have mostly considered for illustrative purposes, focussing on the theory of one abelian Goldstone boson. We could have similarly considered the case of massless vectors or fermions, 
where gauge invariance or supersymmetry mandate derivative interactions.

The constraints are conveniently described by grouping EFTs into two broad classes. The first are  EFTs that emerge from a weakly coupled UV completion, either at tree level or from loops. Here, for $s$ not too much below the physical EFT cut-off, the arcs are reliably approximated by the Wilson coefficients at tree level. Eqs.~(\ref{h1c}-\ref{h4c}) then translate directly into sharp constraints on the Wilson coefficients. In particular, the  energy dependent inhomogeneous constraints dictate the strict convergence of the $s$ expansion of the forward amplitude ${\cal M}(s)$ and rule out the possibility for supersoft EFTs, where ${\cal M}(s)$
grows faster than $s^2$. Indeed  the energy dependent constraints, through their violation, also allow to infer the maximal cut-off of the EFT. As we discussed, even in these weakly coupled EFTs the tree level approximation for the arcs breaks down at sufficiently low $s$ because of mixing at the quantum level with more relevant Wilson coefficients. Consequently in the far IR the allowed region for the running Wilson coefficients differs significantly with respect to the near cut-off region.
This fact is also directly related to the existence of the second class of EFTs,  for which the tree level approximation for the arcs is never realized in their domain of validity. Positivity still implies  strict constraints in couplings space and in particular rules out supersoft EFTs, at least under the simplest assumptions we could check. In this second class of theories the homogeneous constraints, now energy dependent because of quantum effects, can control the maximal allowed UV cut-off.

Our study mostly concerned  the forward amplitude, {but in  section \ref{sec:bf} we extend it to $t\not = 0$. 
We studied the arc first $t$-derivative; in derivatively coupled theories, this too is free of IR-divergences. The bounds we obtain complete our analysis into the realm of theories with suppressed forward amplitude. In particular they show that ${\cal M}(s)$ can be dominated over a limited range of energies by $\propto s^2t$ behaviour, but that anything softer is forbidden. More bounds than those presented here can be derived using the explicit form of Legendre polynomials (in addition to positivity of their derivatives)~\cite{NimaHuangTalks}, but always involve at least two $t$-derivatives. These involve   quantities that are IR divergent in the $m\to 0$ limit, and goes beyond the scope of the present study (see  comments further down).}

Our investigation could be furthered in a number of ways.
One could be to try and connect to the S-matrix bootstrap \cite{Paulos:2017fhb,Guerrieri:2019rwp,Correia:2020xtr}.
The latter approach exploits the full $2\rightarrow 2$ unitarity equation, schematically  $2\text{Im}T> |T|^2$, while our analytical bounds purely exploit positivity $\text{Im}T>0$. Besides trying to implement full unitarity one could perhaps  use an ansatz for the $2\rightarrow 2$ amplitude  similar to  the S-matrix bootstrap approach of \cite{Paulos:2017fhb}.

An obvious way to  extend  our work would be to more systematically study other  instances of derivatively coupled theories. In particular one could consider cases  involving states of different helicity and with a  flavor structure. Here  it would be interesting to consider the forward amplitude for superpositions of helicity and flavor.  A less obvious one would be to consider non derivatively coupled EFTs where the cross section is affected by collinear divergences. In this case the bounds on the Wilson coefficients at some scale $s$  will seemingly have a  dependence, to be determined, on the mass $m$ providing  the IR regulation. Alternatively one could treat the mass $m$ itself as the  RG scale and work with $s\sim m^2$.

\subsection*{Acknowledgements}
We thank Nima Arkani-Hamed, Thomas Becher, Gabriel Cuomo, Brian Henning, Yu-Tin Huang, Paolo Nason and Andrea Wulzer for useful discussions, and Alberto Nicolis, Joao Penedones, Francesco Sgarlata for comments on the manuscript.
The work of MR, FR and RR is supported by the Swiss National Science Foundation under grants no. PP00P2-170578, no. 200021-178999, no. 200020-169696 and through the National Center of Competence in Research SwissMAP. 
BB and FR were also supported by the Munich Institute for Astro- and Particle Physics (MIAPP) which is funded by the Deutsche Forschungsgemeinschaft (DFG, German Research Foundation) under Germany Excellence Strategy D EXC-2094 D 390783311.
JEM, RR and FR  thank the Simons Collaboration on the Non-perturbative Bootstrap for hospitality in ICTP-SAIFR.

\appendix

\section{Analytic Results at Finite-$t$}
\label{sec:appendixt}

It is convenient to rearrange the Hankel matrices $\left(\haf{\ell}{N} \right)$ defined above \eq{posst},
 in terms of blocks of either half-integer or integer arcs, by extending the definition of section~\ref{pa} to 
$\left(\ha{\ell}{N}\right)_{ij}=a_{i+j+\ell}$ where now $\ell$ and $N$ can be  half-integer or integer,  and  $i,j=0,1,\ldots, \lfloor (N-\ell)/2 \rfloor$,  e.g. 
\be
\ha{1/2}{5/2}= \left( 
\begin{array}{cc} a_{1/2} &a_{3/2}  \\
a_{3/2}  & a_{5/2} 
\end{array}
\right)  \, .\label{poss}
\ee
In this way, for $N$ integer, the constraints Eqs.~(\ref{h1c}-\ref{h4c}) on $\haf{\ell}{N}$ (involving both integer and half-integer arcs) can be written in terms of constraints on $\ha{\ell^\prime}{N}$ (integer arcs only for $\ell^\prime$ integer) and  $\ha{\ell^\prime+1/2}{N}$ (half-integer arcs only): 
 \begin{gather}
  \ha{1/2}{} \succ 0\,,  \qquad 
\ha{1}{} \succ\ha{1/2}{}(\ha{0}{})^{-1}\ha{1/2}{}\,,  \nn \\
\ha{3/2}{} \succ\ha{1}{}(\ha{1/2}{})^{-1}\ha{1}{}  \nn\\
\Delta \ha{0}{} \succ 0\,,\qquad \Delta \ha{1/2}{} \succ 0\,, \nn \\
\Delta \ha{1}{} \succ\Delta \ha{1/2}{} (\Delta \ha{0}{})^{-1} \Delta \ha{1/2}{}\,,  \nn \\ 
\Delta \ha{3/2}{} \succ \Delta \ha{1}{} (\Delta \ha{1/2}{})^{-1}\Delta \ha{1}{}  \,,  \label{h1cn} 
\end{gather}
where we defined $\Delta \ha{\ell}{N}\equiv \ha{\ell}{N-1/2}-\bs\ha{\ell+1/2}{N}$ and used  Schur's complement. The lower label  of Hankel matrices is implicit  and corresponds to $N$.

All half-integer arcs   in these conditions are replaced using \eq{dta},
\begin{equation}
\arc_{n+1/2}=\frac{2}{2n+3}\left(a^{(1)}_n-\partial_t \arc_n\right)\,.
\end{equation}
Together with the positivity bounds for $a^{(1)}_n$, involving $a^{(1)}_n$  up to $\lfloor N-1/2  \rfloor$, they define the space of allowed arc derivatives, once we eliminate all the $a^{(1)}_n$.

\bibliography{bibs} 

\end{document}